\documentclass[sigconf, nonacm]{acmart}

\usepackage{subcaption}
\usepackage{balance}
\usepackage{colortbl}
\usepackage{pdfrender}
\newcommand*{\boldcheckmark}{%
  \textpdfrender{
    TextRenderingMode=FillStroke,
    LineWidth=.5pt,
  }{\checkmark}%
}

\AtBeginDocument{%
  }

\renewcommand\footnotetextcopyrightpermission[1]{}
\begin{document}

\title[Your Battery Is a Blast! Safeguarding Against Counterfeit Batteries with Authentication]{Your Battery Is a Blast!\\Safeguarding Against Counterfeit Batteries with Authentication}

\author{Francesco Marchiori}
\orcid{0000-0001-5282-0965}
\affiliation{%
 \institution{University of Padova}
 \streetaddress{Via Trieste, 63}
 \city{Padua}
 \country{Italy}}
\email{francesco.marchiori@math.unipd.it}

\author{Mauro Conti}
\orcid{0000-0002-3612-1934}
\affiliation{%
 \institution{University of Padova}
 \streetaddress{Via Trieste, 63}
 \city{Padua}
 \country{Italy}}
\email{mauro.conti@unipd.it}

\begin{abstract}
Lithium-ion (Li-ion) batteries are the primary power source in various applications due to their high energy and power density.
Their market was estimated to be up to 48 billion U.S. dollars in 2022.
However, the widespread adoption of Li-ion batteries has resulted in counterfeit cell production, which can pose safety hazards to users.
Counterfeit cells can cause explosions or fires, and their prevalence in the market makes it difficult for users to detect fake cells.
Indeed, current battery authentication methods can be susceptible to advanced counterfeiting techniques and are often not adaptable to various cells and systems.
\par
In this paper, we improve the state of the art on battery authentication by proposing two novel methodologies, \textbf{DCAuth} and \textbf{EISthentication}, which leverage the internal characteristics of each cell through Machine Learning models.
Our methods automatically authenticate lithium-ion battery models and architectures using data from their regular usage without the need for any external device.
They are also resilient to the most common and critical counterfeit practices and can scale to several batteries and devices.
To evaluate the effectiveness of our proposed methodologies, we analyze time-series data from a total of 20 datasets that we have processed to extract meaningful features for our analysis.
Our methods achieve high accuracy in battery authentication for both architectures (up to 0.99) and models (up to 0.96).
Moreover, our methods offer comparable identification performances.
By using our proposed methodologies, manufacturers can ensure that devices only use legitimate batteries, guaranteeing the operational state of any system and safety measures for the users.
\end{abstract}



\maketitle

\section{Introduction}
\label{sec:introduction}

Lithium-ion (Li-ion) batteries~\cite{nazri2008lithium} are currently dominating the market due to their higher energy efficiency and low memory effects~\cite{sasaki2013memory}.
An active line of research backs up this growth, with many works on the improvement of their materials and architectures~\cite{scrosati2010lithium}, studies on their aging behaviour~\cite{barre2013review} and devices for their management~\cite{waag2014critical}.
Their rapid development over the years has made them particularly effective solutions in many applications, ranging from small electronic devices to electric vehicles (EVs)~\cite{marinaro2020bringing}.
Indeed, the global market for Li-ion batteries in 2022 was estimated to be up to 48 billion U.S. dollars, with an expected compound annual growth rate of 18.1\% from 2022 to 2030~\cite{grandviewresearch2022lithium}.

\begin{figure*}[!htpb]
    \centering
    \begin{subfigure}{0.33\textwidth}
      \centering
      \includegraphics[width=\textwidth]{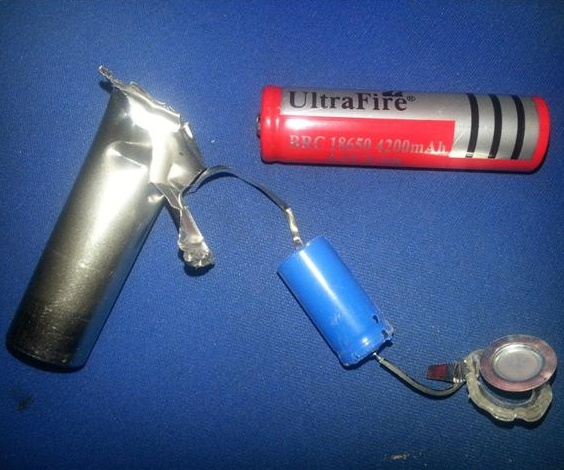}
      \caption{A 18350 cell posing as a 18650 cell.}
      \label{subfig:rewrapping-1}
    \end{subfigure}
    \begin{subfigure}{0.33\textwidth}
      \centering
      \includegraphics[width=\textwidth]{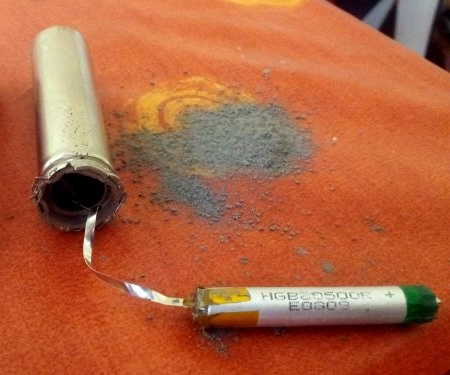}
      \caption{Smaller cell rewrapped with sand.}
      \label{subfig:rewrapping-2}
    \end{subfigure}
    \begin{subfigure}{0.33\textwidth}
      \centering
      \includegraphics[width=\textwidth]{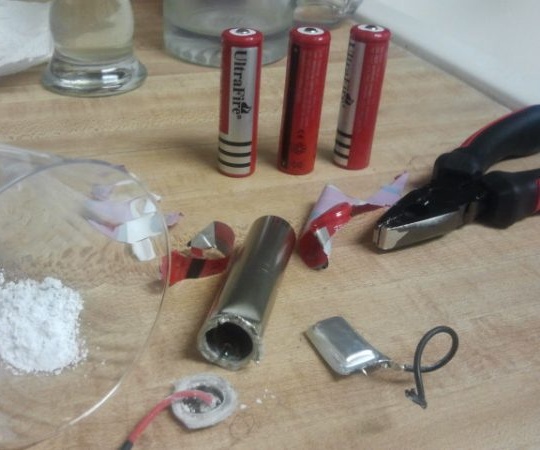}
      \caption{Countefeit containing a small pouch battery.}
      \label{subfig:rewrapping-3}
    \end{subfigure}
    \caption{Examples of counterfeit 18650 batteries.}
    \label{fig:rewrapping}
\end{figure*}

This widespread demand for Li-ion cells, however, is causing a rise in their counterfeiting, i.e., the production of fake or unauthorized replicas of legitimate batteries.
Indeed, counterfeit batteries are flooding the market and the supply chain from the Original Equipment Manufacturer (OEM) to the vendors~\cite{kong2022distribution}.
A significant number of batteries are seized worldwide each year for an estimated value of several million dollars~\cite{obrien2008lithium, us2016cbp}.
These cells are branded similarly to their legitimate counterpart and advertised to work in the same way.
However, they are often made of lower quality materials and have no safety certification.
For these reasons, the risk of fire hazards is significantly higher with respect to the original battery manufacturer cells.
Indeed, the operative temperature for cells can reach critical values, and illegally manufactured batteries or recycled ones might not correctly enforce safety measures.
However, detecting these cells is particularly difficult since their physical condition is often as good as a legitimate one.
By rewrapping the cells, it is possible to fake their rating or capacity, remark them as a different model or even disguise them as a different cell~\cite{saxena2018exploding}.
In Figure~\ref{fig:rewrapping}, we show three examples of counterfeit rewrapped batteries.
In those images, smaller capacity cells are connected through a step-up circuit to the positive and negative terminals of the outer cell.
While cells such as the one in Figure~\ref{subfig:rewrapping-1} might behave similarly to their legitimate counterpart, their lifespan are dramatically decreased.
However, weighting those cells can detect these counterfeiting attempts since they will be much lighter.
Instead, in Figure~\ref{subfig:rewrapping-2} and~\ref{subfig:rewrapping-3}, the size gap is filled with powders or other materials to compensate for the weight difference, making the detection of those fakes much harder.
Users generally report many issues or power interruptions during their usage.\footnote{\url{http://e-motion.lt/2016/04/13/ultrafire-5800mah/}}
Others that use those counterfeited cells in on-road vehicles instead are particularly in danger and can also pose a threat to nearby vehicles.\footnote{\url{https://endless-sphere.com/forums/viewtopic.php?t=80451}}

\paragraph{Contribution}
In this paper, we present \textbf{DCAuth} and \textbf{EISthentication}, two novel solutions for the automatic authentication and identification of Li-ion battery cells.
While other methods for the authentication of Li-ion batteries are present in the literature, to the best of our knowledge, ours are the first to consider only physical and chemical features and thus do not rely on external devices or Challenge-Response (CR) authentication protocols.
This is significant since other methods have been proven weak to several attacks or not scalable to many different battery models.
Table~\ref{tab:comparison} shows an overview of the strengths of DCAuth and EISthentication compared to other popular authentication methods.
Our methodologies take advantage of data retrieved through, respectively, Differential Capacity Analysis (DCA) and Electrochemical Impedance Spectroscopy (EIS).
The former is a technique used in electrochemical measurements to study the behavior of the electrode-electrolyte interface.
The latter is a non-destructive method for characterizing Li-ion batteries~\cite{meddings2020application}.
By considering different cell models and performing DCA or EIS on them at different States Of Charge (SOC) and States Of Health (SOH), we extract features to train several Machine Learning (ML) models.
Of the different models considered, the best results are reached by the Random Forest (RF) classifier, which obtains scores up to 0.96 in the authentication of different battery samples.
The main contributions of our work can be summarized as follows:
\begin{itemize}
    \item We improve the state of the art on battery authentication by leveraging the internal characteristics of each cell and Machine Learning models.
    We propose two methodologies (\textbf{DCAuth} and \textbf{EISthentication}) for the authentication and identification of batteries in any setting or environment.
    \item We define and publish a common procedure to process battery data coming from different datasets.
    Our procedure considers several types of equipment and their characteristics to facilitate future research in this field.
    \item We evaluate our methodologies on 20 datasets we collected and processed.
    Our evaluation is differentiated into various steps and considers both the authentication and identification tasks.
    \item We make available the code and implementation for all of our methodologies, including both the Machine Learning models and their experimental evaluation.
    Our repositories can be accessed at \url{https://github.com/Mhackiori/DCAuth} and \url{https://github.com/Mhackiori/EISthentication}.
\end{itemize}

\def\anlge{61}

\begin{table}[!htpb]
  \centering
  \caption{Comparison of different methodologies for battery authentication. In each cell, we put a checkmark if the considered methodology is resilient against a particular attack or weakness.}
  \label{tab:comparison}
  \begin{tabular}{lcccc}
    \toprule
    \textbf{Method} & \rotatebox{\anlge}{\textbf{Cloning}} & \rotatebox{\anlge}{\textbf{Replay Attacks}} & \rotatebox{\anlge}{\textbf{Unscalability}} & \rotatebox{\anlge}{\textbf{Rewrapping}} \\
    \midrule
    Markings &  & \checkmark & \checkmark &  \\
    External Features &  & \checkmark & \checkmark &  \\
    Form Factor &  & \checkmark & \checkmark &  \\
    Resistor &  & \checkmark & \checkmark &  \\
    Chip & \checkmark &  &  &  \\
    CR (in clear) & \checkmark &  &  &  \\
    CR (encrypted) & \checkmark & \checkmark &  &  \\
    \textbf{DCAuth} & \boldcheckmark & \boldcheckmark & \boldcheckmark & \boldcheckmark \\
    \textbf{EISthentication} & \boldcheckmark & \boldcheckmark & \boldcheckmark & \boldcheckmark \\
    \bottomrule
  \end{tabular}
\end{table}

\paragraph{Organization}
The paper is organized as follows.
Section~\ref{sec:relatedworks} reviews related works on different battery authentication methods and the techniques used to perform DCA and EIS.
An overview of the system model and examples of practical deployment are given in Section~{\ref{sec:systemmodel}}.
In Section~\ref{sec:datasets}, we give an overview of the data collection process and the characteristics of the different datasets used for each methodology.
In Section~\ref{sec:methodologies}, we explain in detail our methodology, presenting an experimental evaluation in Section~\ref{sec:results}.
Section~{\ref{sec:discussion}} presents a discussion of the obtained results and possible limitations of our methodologies.
Finally, Section~\ref{sec:conclusions} concludes this work.
\section{Related Works}
\label{sec:relatedworks}

We now overview several techniques that are currently used for battery authentication and their flaws that led to the need for a device-independent authentication method (Section~\ref{subsec:batteryauthentication}).
Furthermore, we summarize the core concepts behind the techniques that we use in our methodologies: Differential Capacity Analysis (Section~\ref{subsec:dca}) and Electrochemical Impedance Spectroscopy (Section~\ref{subsec:eis}).

\subsection{Battery Authentication}
\label{subsec:batteryauthentication}

Battery authentication refers to the process of verifying the authenticity of a battery, typically to ensure that it is a genuine product and not a counterfeit or a lower-quality substitute.
Several methods can be used to authenticate batteries, including visual inspection, chemical analysis, and various non-destructive testing techniques.
In this review, we will only analyze the methods that authenticate the battery before a system is allowed to function from it.
Indeed, the system should refuse power from a battery that failed the authentication process~\cite{tsao2018how}.

\begin{itemize}

\item \textbf{Visual Inspection} --
One method of battery authentication is visual inspection, which involves looking for physical characteristics that are unique to the manufacturer or product.
For example, many batteries have specific markings or labels that can be used to identify them.
In addition, the appearance of the battery itself, such as its color, shape, and size, can also be used to verify its authenticity.
This method, however, is vulnerable to rewrapping, and in general, it is really easy to replicate the markings on the wrapping or other external characteristics.

\item \textbf{Form Factor} --
A similar authentication method involves leveraging the form factor of the battery casing and connectors.
However, third parties can easily reproduce the physical properties of the cells to create their counterfeit samples.

\item \textbf{Resistor or Chip} --
Another basic approach is to place a resistor in the battery pack to identify its type.
Alternatively, instead of a resistor, it is possible to place a memory chip containing several data on the battery, including type, ID, and manufacturing date.
However, these resistors or chips can be extracted and replaced on counterfeit samples, which will pose as legitimate to a prover.

\item \textbf{Challenge-Response Protocols} --
It is possible to use a Challenge-Response (CR) protocol between the chip and the prover to authenticate the battery.
In EVs, this is usually handled by the fuel gauges, which manage data coming from the battery to monitor its State Of Function (SOF)~\cite{rezal2014orion}.
The most basic implementation of this protocol is using an unchanging stream of bits that, however, can be sniffed by an attacker and thus is vulnerable to replay attacks.
More advanced gauges instead include some cryptographic hash function in the protocol, making it impossible for an attacker to steal the authentication codes~\cite{al2019sha}.
While this method is particularly effective in preventing cloning, it requires the creation of the keys during the battery's manufacturing process.
This makes the legitimate battery sample compatible only with a specific family of fuel gauges.

\end{itemize}

We thus propose two techniques for battery authentication that do not depend on external devices and instead leverage physical and chemical characteristics to determine the legitimacy of a battery.
Moreover, updating the labels corresponding to the legitimate samples makes it possible to contemplate the legitimacy of new battery packs or architectures that did not exist at the time of the implementation.

\subsection{Differential Capacity Analysis}
\label{subsec:dca}

Differential Capacity Analysis is a common method for examining the interfaces between electrodes and electrolytes in electrochemical systems.
Differential Capacity can track the increase in capacity when charged or a decrease when discharged of an electrochemical system.
By creating a plot of differential capacity versus voltage, a ``fingerprint'' of the system can be obtained and monitored over time.
This provides insight into the system's thermodynamics and kinetics as the characteristic features of the curve change.

While initially explored as a quality control measure, differential capacity analysis can aid in foreseeing malfunctions by observing variations in the electrochemical characteristics of a system~\cite{zhao2022data}.
This capacity for prediction would aid in preventing safety hazards and ensuring that a battery or cell is retired only when it can no longer function effectively.

The use of DCA to estimate State Of Charge and State Of Health has also been widely researched~\cite{torai2016state, kurzweil2022differential}. Battery management and monitoring rely on SOH estimation to gain insights into a battery's condition over time.
Since DCA works by analyzing the difference in capacity between the initial discharge and subsequent cyclic discharge of a battery, by measuring this difference, it is possible to estimate the amount of irreversible capacity loss.
Using DCA, battery manufacturers and designers can enhance their understanding of the battery's performance throughout its lifespan, improving battery longevity and reducing overall system costs.

With DCAuth, we leverage the relationship between DCA and the degradation of battery samples to authenticate and identify different models and architectures.
Indeed, some reaction processes consume electrolyte species and/or active lithium in Li-ion cells.
These are often referred to as parasitic reactions.
These electrochemical reactions occur at the interfaces between the high-voltage positive electrode or low-voltage negative electrode and the electrolyte.
Since parasitic reactions have been shown to alter peaks in the differential capacity plot~\cite{krupp2020incremental}, our model can extract meaningful features and use them to correctly identify and authenticate the model and architecture of battery samples.
Furthermore, since both voltage and capacity measurements are fairly accessible in most systems, DCA estimations can be performed on many different devices~\cite{calearo2022agnostic}.
This makes DCAuth particularly affordable and inexpensive in most authentication scenarios.

\subsection{Electrochemical Impedance Spectroscopy}
\label{subsec:eis}

Electrochemical Impedance Spectroscopy (EIS) is a powerful analytical technique that allows for the characterization of electrochemical systems by measuring their electrical impedance~\cite{chang2010electrochemical}.
This technique is widely used in various fields, including materials science, corrosion science, electrochemistry, and biomedical engineering, as it provides valuable insights into the behavior and properties of a system~\cite{lasia2002electrochemical}.
In an EIS measurement, a small AC voltage is applied to an electrochemical cell, and the resulting current is measured.
The system's impedance is then calculated as the ratio of the applied voltage to the measured current.
It is possible to obtain a detailed picture of the system's electrical properties by measuring the impedance over a range of frequencies.
In addition to its analytical capabilities, EIS has several other advantages.
It is a non-invasive technique, meaning it does not require the destruction of the sample or the addition of chemical probes.
It is also relatively simple to set up and perform and can be carried out at a wide range of temperatures~\cite{buscaglia2021roadmap}.

\begin{figure*}[!htpb]
  \centering
  \begin{subfigure}{0.33\textwidth}
     \centering
     \includegraphics[width=\textwidth]{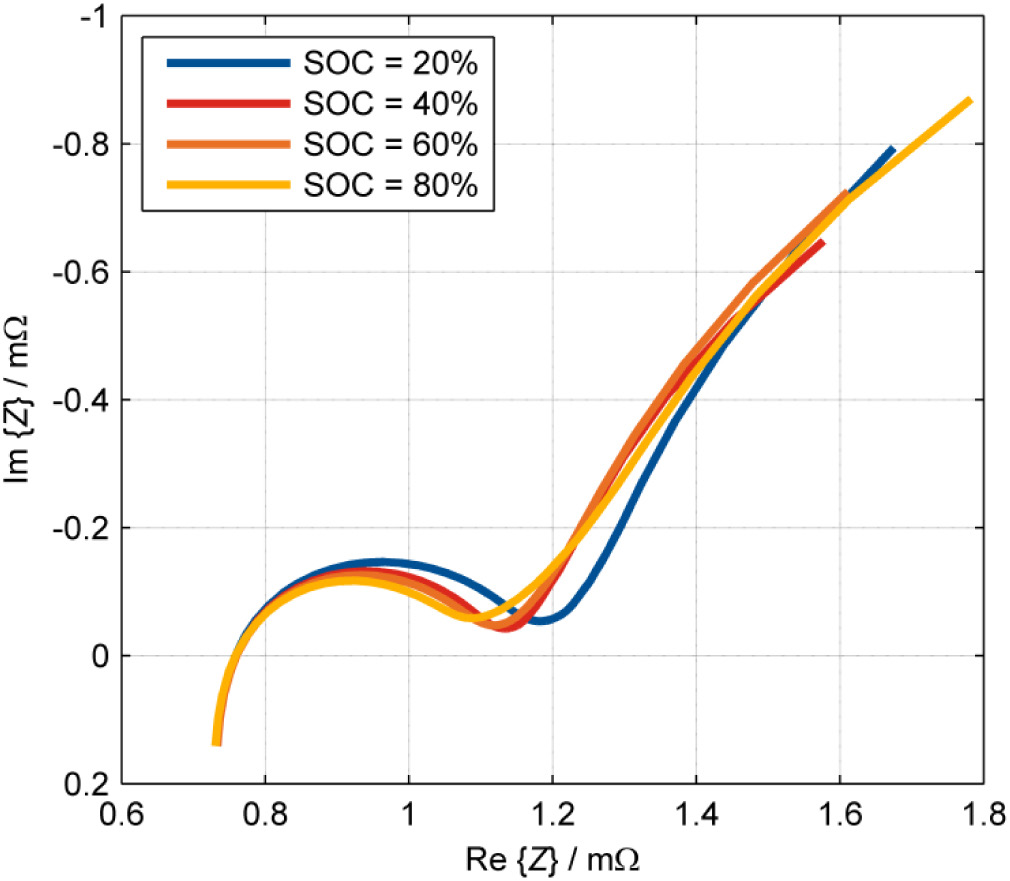}
     \caption{Dependance on SOC.}
     \label{subfig:soc}
  \end{subfigure}
  \begin{subfigure}{0.33\textwidth}
     \centering
     \includegraphics[width=\textwidth]{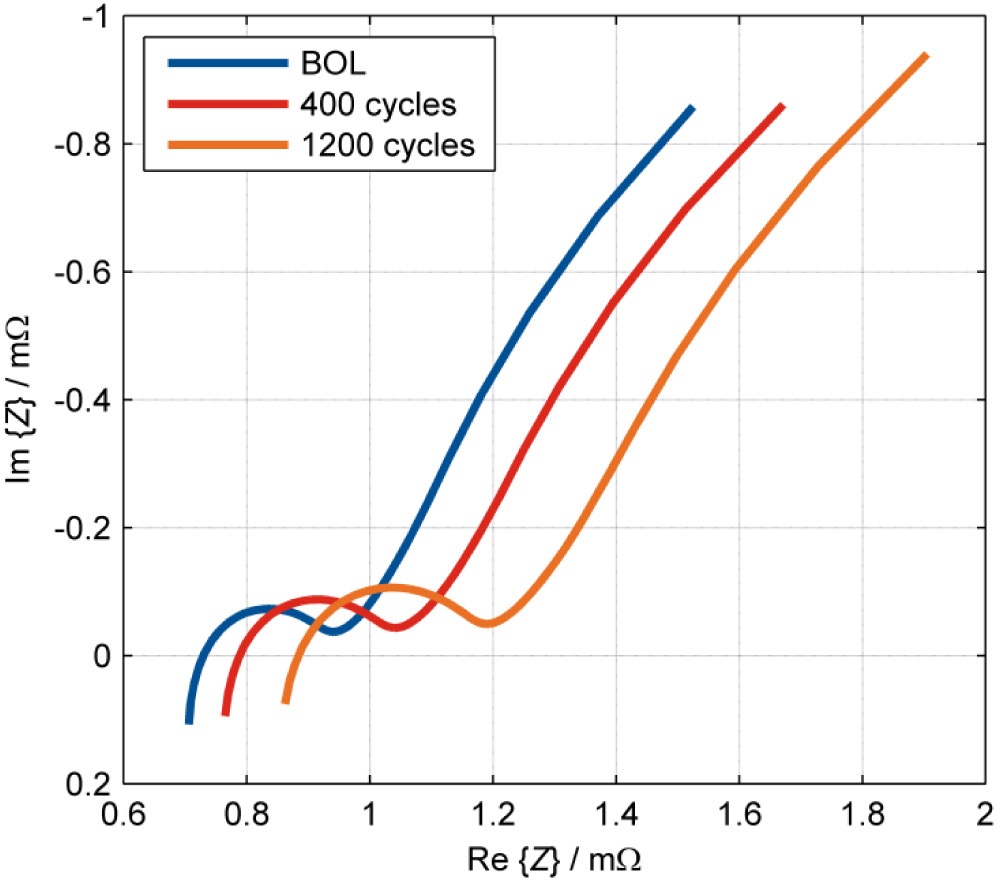}
     \caption{Dependance on SOH.}
     \label{subfig:temp}
  \end{subfigure}
  \begin{subfigure}{0.33\textwidth}
     \centering
     \includegraphics[width=\textwidth]{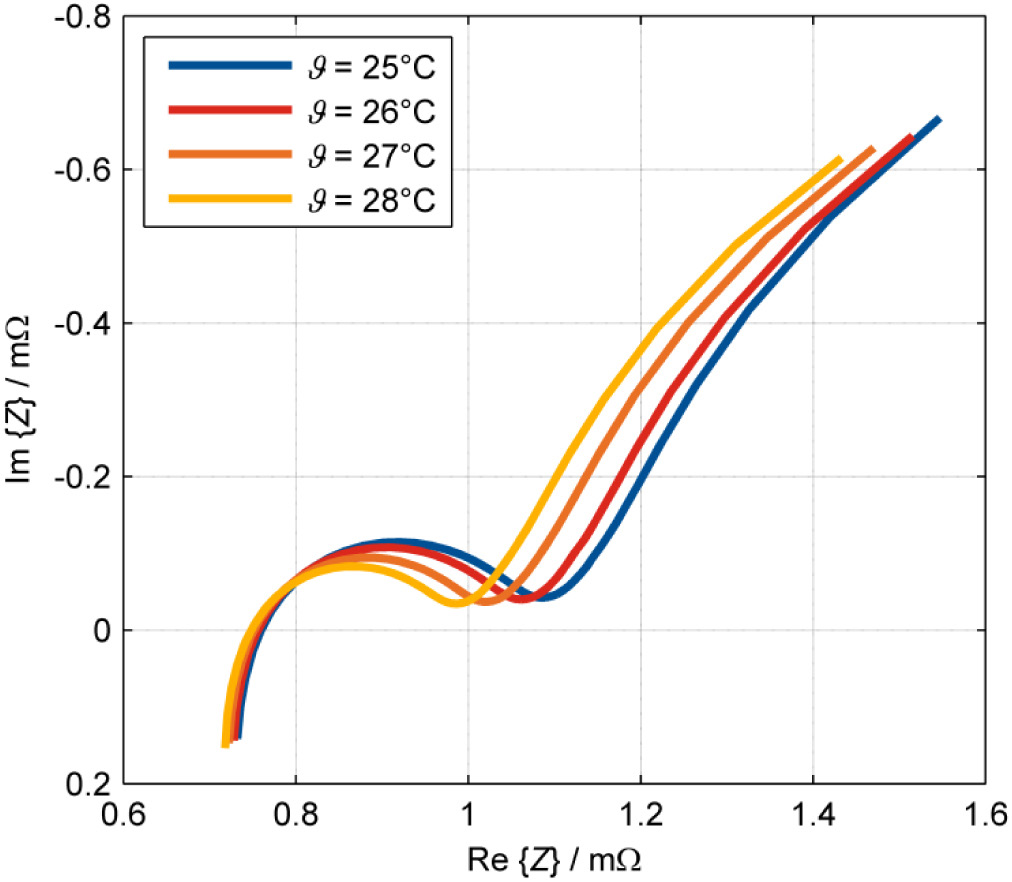}
     \caption{Dependance on temperature.}
     \label{subfig:soh}
  \end{subfigure}
  \caption{Nyquist plots of EIS for different parameters. Images from~\cite{lohmann2015employing}.}
  \label{fig:eis}
\end{figure*}

EIS can be used to study the electrical and electrochemical properties of batteries and supercapacitors, such as the charge/discharge behavior, the rate capability, and the cycling stability.
It can provide valuable information about the performance and durability of these energy storage devices.
Especially with lithium-ion batteries, several studies have been performed on the usage of EIS for the estimation of the State Of Charge (SOC)~\cite{ran2010prediction, westerhoff2016electrochemical}, State Of Health (SOH)~\cite{li2022electrochemical, jiang2022comparative} and other relevant data for the diagnosis of the cells~\cite{zhuang2012diagnosis}.
In particular, SOC, SOH, and temperature are important parameters that can influence materials and devices' electrical and electrochemical properties.
They can therefore affect the Nyquist plot of the impedance.
The Nyquist plot is retrieved by plotting the imaginary part of the impedance extracted through EIS versus its real part~\cite{lohmann2015employing}.
Figure~\ref{fig:eis} gives a graphical representation of this influence.

Given the unique trend of the impedance with respect to the electrochemical components of the battery, with EISthentication we leverage the real and imaginary parts of the plot to extract features that will be used to train Machine Learning classifiers.
These features allow us to profile each battery sample and thus authenticate different models of Li-ion batteries.
\section{System Model}
\label{sec:systemmodel}

\begin{figure}[!htpb]
    \centering
    \includegraphics[width=\linewidth]{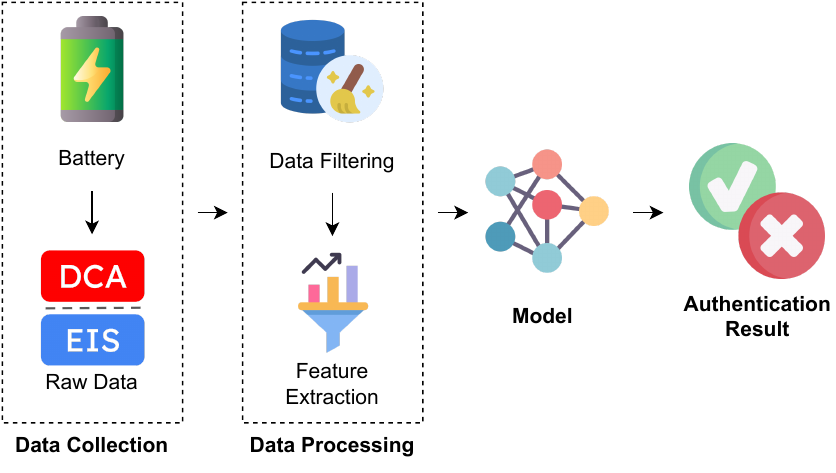}
    \caption{System Model of DCAuth and EISthentication.}
    \label{fig:flow}
\end{figure}

In this section, we outline our conceptual framework and provide possible implementations of our authentication system.
Figure~{\ref{fig:flow}} shows an overview of the system model.
While the majority of the detailed steps are shared between DCAuth and EISthentication, their difference lies primarily in the data collection process, which will be treated separately in Section~{\ref{subsec:dcasm}} and Section~{\ref{subsec:eissm}}.

\subsection{DCAuth System Model}
\label{subsec:dcasm}

{As described in Section~{\ref{subsec:dca}}, Differential Capacity Analysis is a technique that requires only voltage and capacity measurements in order to be performed.
These data types are accessible in most battery-powered systems since they are often already used for SOC or SOH estimation.
Thus, DCAuth can be performed by the device without external equipment.
To make the process viable for compact, portable devices, authentication is carried out using Machine Learning models of minimal complexity, adding minimal overhead to the system.
Essentially, the authentication system can constitute a software module of the system.
This approach allows for pre-executed model training, while Over-The-Air (OTA) updates enable parameter adjustments to enhance performance or incorporate new battery types.
During the regular usage of the device, voltage and capacity data are collected and processed to estimate the DCA plot over one battery cycle.
This cycle can be either the charging or discharging cycle, or both can be considered to increase accuracy.
After sufficient data has been gathered, a filtering process is employed to eliminate any artifacts, and features are extracted from the processed plots.
Finally, the data samples are tested on the deployed model, and an authentication response is generated.
While authentication can be performed once each time a battery swap is detected, it is also possible to provide continuous authentication while concurrently gathering data to enhance performance.

\subsection{EISthentication System Model}
\label{subsec:eissm}

The main difference of using EIS instead of DCA relies on the data collection process.
While with DCAuth it is possible to gather data for authentication while the device is in use, with EISthentication external equipment is needed in order to compute the real and imaginary part of the battery impedance.
However, since EIS can be performed at any stage of the battery's life and charge level, data collection takes significantly less time than DCA, depending on the battery implementation.
This allows authentication to be performed before even powering the device.
In this way, the system model can be fully implemented in an external device comprising both the measurement apparatus and the computational power necessary for data processing and testing on the models.
Furthermore, several research works have studied the possibility of retrieving EIS without specialized equipment, which in the future might close the gap between the convenience of the internal system model of DCAuth and the quickness of EISthentication~{\cite{troltzsch2006characterizing, howey2013online, lohmann2015employing}}.
\section{Datasets}
\label{sec:datasets}

This section details the used datasets and how they have been integrated with our methodologies.
We first detail our data collection process (Section~\ref{subsec:collection}) while focusing separately on the datasets for DCA (Section~\ref{subsub:dcadata}) and the ones for EIS (Section~\ref{subsub:eisdata}).
After that, we describe the post-processing performed on these datasets to remove artifacts and clean the data (Section~\ref{subsec:processing}).

\subsection{Collection}
\label{subsec:collection}

One of the most crucial aspects of our work is the data collection process.
Indeed, by using Machine Learning models to perform our tasks, we need huge amounts of data that fulfill specific requirements.

\begin{itemize}
    \item \textbf{Different battery models} -- 
    Since our task can be summarized as a classification task, we need as many classes as possible to represent a real-world application for our tool accurately.
    However, those classes' types depend on what we want to classify.
    We can indeed classify battery models, i.e., a specific production line by the manufacturer, or battery architectures, i.e., the physical and chemical design and structure of the battery (e.g., Lithium Iron Phosphate, Nickel Manganese Cobalt).
    Thus, we need datasets that contain data on many different battery cells, which include different form factors, lithium-ion battery designs, and nominal voltage and capacity values.
    \item \textbf{Varying conditions} --
    Both our methodologies are designed to be adaptable to many different environments (e.g., authentication of a battery pack in an EV, authentication of a battery in a portable device).
    Therefore, our datasets should include data on battery cells that have been cycled under different conditions.
    In particular, we highlighted how SOC, SOH, and temperature can influence DCA and EIS plots.
    Therefore, our study should consider the different stages of each battery lifecycle.
    \item \textbf{Cycling type} --
    Data itself is extracted from each battery sample by stressing it through many charge cycles, i.e., by charging and discharging it.
    However, the way in which each battery is cycled can vary depending on different conditions.
    For example, a typical method of charging small batteries is Constant Current Constant Voltage (CCCV), where the cell is initially charged at a constant current, but when it is nearly full, it switches to constant voltage.
    These constant values, however, can change depending on the battery model and its future implementation.
    Furthermore, many different discharging profiles are publicly available to mimic their implementation on different environments (e.g., driving profiles~\cite{smith2011characterization}).
    Therefore, our datasets should include heterogeneous data gathered from different cycling types and profiles. 
\end{itemize}

These requirements impose some heavy constraints on our data creation process.
For this reason, we contacted many different private companies to collaborate with them on the generation of the dataset.
To our dismay, no institution agreed to share their data with us or to generate new datasets.
This is due to several reasons.
First, cycling many different batteries is an expensive procedure requiring constant maintenance.
On top of the energy costs in maintaining the infrastructure, cell samples can be quite costly when we consider scenarios like automotive, where an EV battery can cost from \$5000 to \$20000.
Another aspect to consider is the degradation of those cells.
Indeed, by constantly cycling them, they would degrade much faster than with regular usage.
For these reasons, we instead collect most of the available datasets in the literature and process them to extract the data that we need for each of the methodologies.

The usage of publicly available datasets restricts us to using only original battery sample data in their intended application.
However, this aspect does not prevent us from detecting counterfeit batteries for the following reasons.
\begin{itemize}
    \item As shown in Figure~{\ref{fig:rewrapping}}, counterfeit batteries are obtained by masquerading a lower-quality battery as a different cell.
    This implies that recognizing a fake sample is the same as determining whether the utilized battery is appropriate for that specific purpose.
    Therefore, the following sections will treat different battery models as counterfeit based on their legitimate application.
    \item While pursuing variety in the datasets regarding condition parameters and type of cycling performed, we obtain data from different cells and different environments.
    This allows us to gather information on different implementations of the same cell and study similar cycling procedures' effects on different batteries.
    Thus, achieving good identification and authentication results will indicate the ability of our methodologies to distinguish between different battery usages and thus detect the legitimacy of a battery in a specific application. 
\end{itemize}

\subsubsection{DCA Datasets}
\label{subsub:dcadata}

As stated in Section~\ref{subsec:dca}, to perform Differential Capacity Analysis, we need measurements on the voltage and capacity of each battery sample.
These measurements are quite common in many different datasets since they are fairly accessible and easy to retrieve.
To collect the most amount of data available in the literature, we refer to a survey performed by Dos Reis et~al.~\cite{dos2021lithium}.
In this work, the authors analyze over 30 datasets and their characteristics.
This allows us to choose among them the ones that are most in line with our requirements.
Indeed, while most of the listed datasets can be eligible for our work, some are now unavailable or do not contain data needed for processing.
In Table~\ref{tab:dca}, we overview each dataset we use for DCAuth.
Furthermore, we highlight many different aspects of each dataset, such as cycling type, cycling equipment, and other data that can be retrieved from their respective papers.
Indeed, for our purposes, having datasets in which batteries have been cycled with different techniques can help us identify and authenticate each cell under different stressing conditions.
Moreover, by being differential, DCA is less sensitive to the absolute values of the charge and voltage than other techniques.
This is because the differential value is calculated as a ratio of changes in capacity and voltage rather than absolute values.
Thus, we ensure that the type of cycling used in the datasets does not introduce bias in our models' performance.

As we notice, data in each dataset has been extracted with different equipment and in many different formats.
Thus, as a preprocessing procedure, we manage the different file formats by converting each dataset into many \texttt{csv} files containing only the data that we need (i.e., voltage and capacity).
To allow researchers to work in this field more efficiently, we publish both the code for our preprocessing procedure and the processed datasets in our repository.

\def\onelines{16pt}
\def\twolines{24pt}
\def\threelines{36pt}
\def\batmodlen{2.25cm}
\def\batlen{2.875cm}
\def\eqlen{2cm}
\def\datalen{5.5cm}

\begin{table*}[!htpb]
  \centering
  \caption{Datasets used for DCAuth.}
  \label{tab:dca}
  \begin{tabular}{l|p{\batmodlen}|p{\batlen}|p{\eqlen}|p{\datalen}}
    \toprule
    \textbf{Dataset} & \textbf{Battery Model} & \textbf{Battery Architecture} & \textbf{Equipment} & \textbf{Data} \\
    \midrule
        \rowcolor{gray!15}
        Berkley~\cite{berkley}
        &
        \parbox[][\twolines][c]{\batmodlen}{Sanyo 18650}
        &
        \parbox{\batlen}{LCO/Graphite}
        &
        \parbox{\eqlen}{N.A.}
        &
        \parbox{\datalen}{CCCV, MCC, CP-CV, and Boostcharge cycles at various C-rates.}
         \\\hline
         
        CALCE\_1~\cite{zheng2016influence, xing2014state, he2014state}
        &
        \parbox[][\twolines][c]{\batmodlen}{INR 18650-20R}
        &
        \parbox{\batlen}{NMC/Graphite}
        &
        \parbox{\eqlen}{Arbin BT2000}
        &
        \parbox{\datalen}{Low Current and Incremental Current OCV tests, Dynamic Test Profiles.}
         \\\hline
         
        \rowcolor{gray!15}
        CALCE\_2~\cite{xing2014state, he2014state}
        &
        \parbox[][\twolines][c]{\batmodlen}{ANR26650M1A}
        &
        \parbox{\batlen}{LFP}
        &
        \parbox{\eqlen}{Arbin BT2000}
        &
        \parbox{\datalen}{Low Current OCV tests, Dynamic Test Profiles.}
         \\\hline
         
        CALCE\_3~\cite{he2011prognostics, xing2013ensemble, williard2013comparative}
        &
        \parbox[][\twolines][c]{\batmodlen}{CS2}
        &
        \parbox{\batlen}{LCO}
        &
        \parbox{\eqlen}{Arbin BT2000, CADEX Tester}
        &
        \parbox{\datalen}{CCCV with different discharging protocols.}
         \\\hline
         
        \rowcolor{gray!15}
        CALCE\_4~\cite{he2011prognostics, xing2013ensemble}
        &
        \parbox[][\twolines][c]{\batmodlen}{CX2}
        &
        \parbox{\batlen}{LCO}
        &
        \parbox{\eqlen}{Arbin BT2000, CADEX Tester}
        &
        \parbox{\datalen}{CCCV with different discharging protocols.}
         \\\hline
         
        CALCE\_5~\cite{saxena2016cycle}
        &
        \parbox[][\onelines][c]{\batmodlen}{PL Samples}
        &
        \parbox{\batlen}{LCO/Graphite}
        &
        \parbox{\eqlen}{Arbin BT2000}
        &
        \parbox{\datalen}{CCCV cycles on different SOC ranges.}
         \\\hline
         
        \rowcolor{gray!15}
        EVERLASTING\_1~\cite{https://doi.org/10.4121/14377295.v1}
        &
        \parbox[][\twolines][c]{\batmodlen}{INR18650 MJ1}
        &
        \parbox{\batlen}{NMC}
        &
        \parbox{\eqlen}{Maccor}
        &
        \parbox{\datalen}{Aged at different C-rated and temperature within a 10-90\% SOC window.}
         \\\hline
         
        EVERLASTING\_2~\cite{https://doi.org/10.4121/13739296.v1}
        &
        \parbox[][\twolines][c]{\batmodlen}{INR18650 MJ1}
        &
        \parbox{\batlen}{NMC}
        &
        \parbox{\eqlen}{Maccor}
        &
        \parbox{\datalen}{Aged at different C-rated and temperature within a 10-90\% SOC window.}
         \\\hline
         
        \rowcolor{gray!15}
        HNEI~\cite{devie2018intrinsic}
        &
        \parbox[][\twolines][c]{\batmodlen}{ICR18650 C2}
        &
        \parbox{\batlen}{LCO/NMC}
        &
        \parbox{\eqlen}{Arbin}
        &
        \parbox{\datalen}{Cycled at 1.5C to 100\% DOD for more than 1000 cycles at room temperature.}
         \\\hline
         
        OX~\cite{raj2020investigation, birkl2017a}
        &
        \parbox[][\twolines][c]{\batmodlen}{SLPB533459H4}
        &
        \parbox{\batlen}{LCO}
        &
        \parbox{\eqlen}{Maccor 4200}
        &
        \parbox{\datalen}{1-C charge, 1-C discharge, pseudo-OCV charge, pseudo-OCV discharge.}
         \\\hline
         
        \rowcolor{gray!15}
        OX\_1~\cite{reniers2020a}
        &
        \parbox[][\onelines][c]{\batmodlen}{SLPB533459H4}
        &
        \parbox{\batlen}{LCO}
        &
        \parbox{\eqlen}{Maccor 4200}
        &
        \parbox{\datalen}{CCCV charge and CCCV discharge.}
         \\\hline
         
        OX\_2~\cite{raj2020a}
        &
        \parbox[][\twolines][c]{\batmodlen}{NCR18650BD}
        &
        \parbox{\batlen}{NCA}
        &
        \parbox{\eqlen}{Maccor 4200}
        &
        \parbox{\datalen}{Different combined profile groups with reference performance tests.}
         \\\hline
         
        \rowcolor{gray!15}
        SNL~\cite{preger2020degradation}
        &
        \parbox[][\threelines][c]{\batmodlen}{
        $\bullet$ APR18650M1A\\
        $\bullet$ NCR18650B\\
        $\bullet$ LG 18650HG2
        }
        &
        \parbox{\batlen}{
        $\bullet$ LFP\\
        $\bullet$ NCA\\
        $\bullet$ NMC
        }
        &
        \parbox{\eqlen}{Arbin LBT21084}
        &
        \parbox{\datalen}{Charged at 0.5C, discharged at 3C. Cycled at three different SOC ranges (0-100, 20-80, 40-60) at CC or CCCV.}
         \\\hline
         
        TRI\_1~\cite{severson2019data}
        &
        \parbox[][\threelines][c]{\batmodlen}{APR18650M1A}
        &
        \parbox{\batlen}{LFP/Graphite}
        &
        \parbox{\eqlen}{Arbin LBT 48ch}
        &
        \parbox{\datalen}{Batteries charged with a one-step or two-step fast-charging policy depending on SOC.}
         \\\hline
         
        \rowcolor{gray!15}
        TRI\_2~\cite{attia2020closed}
        &
        \parbox[][\twolines][c]{\batmodlen}{APR18650M1A}
        &
        \parbox{\batlen}{LFP/Graphite}
        &
        \parbox{\eqlen}{Arbin LBT 48ch}
        &
        \parbox{\datalen}{Cells are cycles with one of 224 six-step 10-minutes fast charging protocols.}
         \\\hline
         
        UCL~\cite{heenan2020advanced}
        &
        \parbox[][\twolines][c]{\batmodlen}{INR18650 MJ1}
        &
        \parbox{\batlen}{NMC/Graphite}
        &
        \parbox{\eqlen}{Maccor 4200}
        &
        \parbox{\datalen}{CC charging at 1.5 A until 4.2 V. Discharging at 4.0 A to 2.5 V.}
         \\\hline
         
        \rowcolor{gray!15}
        UL-PUR~\cite{juarez2020degradation}
        &
        \parbox[][\twolines][c]{\batmodlen}{NCR18650B}
        &
        \parbox{\batlen}{NCA}
        &
        \parbox{\eqlen}{Arbin BT2543}
        &
        \parbox{\datalen}{Discharged to 2.7 V (CC), charged to 4.2 V (CCCV).}
         \\
    \bottomrule
  \end{tabular}
\end{table*}

\subsubsection{EIS Datasets}
\label{subsub:eisdata}

While datasets containing voltage and capacity can be easily found in the literature, studies on their internal impedance or Electrochemical Impedance Spectroscopy are far more rare.
Indeed, EIS is a complex technique that requires specialized equipment and trained technicians to perform the experiments.
Furthermore, EIS equipment is expensive and requires regular maintenance, which makes it difficult for researchers with limited funding to perform EIS experiments.
However, many research works are moving toward estimating Electrochemical Impedance Spectroscopy in a device or a vehicle.
While some works still need some additional electronics~\cite{troltzsch2006characterizing}, others leverage data on internal components such as the excitation of the motor controller~\cite{howey2013online} or real driving data~\cite{lohmann2015employing}.
However, due to the unavailability of such datasets, our focus remains exclusively on EIS data obtained directly from measurements.

The first dataset that we use is the SiCWell dataset~\cite{goldammer2022impact}.
This dataset contains data on automotive-grade lithium-ion pouch bag cells cycled with two popular driving profiles (sWLTP and UDDS).
The data collection's main focus is investigating the influence of ripple currents in the EV battery.
Still, several checkups on each pouch bag are performed periodically.
In particular, the authors performed electrochemical impedance spectroscopy from 0.001 to 50000 Hz using an EIS-meter for each sample.
Each procedure has been repeated at every checkup for four different SOCs (20\%, 40\%, 60\%, and 80\%).
The processing resulted in a dataset containing 37 different battery samples.
The dataset is freely available on IEEE DataPort\footnote{\url{https://ieee-dataport.org/open-access/sicwell-dataset}}.

To train and evaluate the capability of our methodology in distinguishing different cell architectures, we will use another dataset provided by Sandia National Laboratories (SNL)~\cite{barkholtz2017database}.
In this dataset, authors considered several commercial 18650 Li-ion battery models with four different chemistries: LiCoO$_{\mathrm{2}}$ (LCO), LiFePO$_{\mathrm{4}}$ (LFP), LiNi$_{\mathrm{x}}$Co$_{\mathrm{y}}$Al$_{\mathrm{1-x-y}}$O$_{\mathrm{2}}$ (NCA), and LiNi$_{\mathrm{0.80}}$Mn$_{\mathrm{0.15}}$Co$_{\mathrm{0.05}}$ (NMC).
The cells have been tested by cycling them at different temperatures, and EIS has been performed at the same SOC to guarantee consistent results.
The frequency range for EIS is from 0.1 to 100000 Hz with a 0.010 V perturbation and has been performed at five different temperatures (5°C, 15°C, 25°C, 35°C, and 45°C).
Furthermore, EIS has also been performed on brand-new cells at pristine conditions.
The dataset is freely available in the official Sandia R\&D Data Repository\footnote{\url{https://www.sandia.gov/ess/tools-resources/rd-data-repository}}.

The final dataset that we consider is provided by Zhang et~al. in~\cite{zhang2020identifying}.
In their work, the authors cycled 12 LCO graphite Li-ion cells at three different temperatures (25°C, 35°C, and 45°C) and performed EIS at nine different stages of their charging/discharging cycles from 0.02 to 20000 Hz.
Each cell is cycled until it reaches its End Of Life (EOL), which is defined as when its SOH drops below 80\% its initial value.
The dataset is freely available on Zenodo\footnote{\url{https://zenodo.org/record/3633835}}.

\subsection{Processing}
\label{subsec:processing}

As anticipated in Section~\ref{subsub:dcadata}, collecting different datasets from the literature often comes at the cost of having different data formats and extraction mechanisms.
While the datasets that we use for EISthentication contain the raw spectroscopy values (and thus Nyquist plots could be directly generated), the ones for DCAuth instead contain only the voltage ($V$) and capacity ($Q$) values.
Thus, differential capacity (i.e., $dQ/dV$) must be calculated from those values and processed to remove artifacts.
Indeed, the raw differential capacity can be computed directly with the following equation.
\begin{equation}
\label{eq:dca}
    \left( dQ/dV \right)_i = \frac{\left( Q_i - Q_{i-1} \right)}{\left( V_i - V_{i-1} \right)}.
\end{equation}
However, values obtained in this way often present some noise and contain many bogus local maxima/minima, as shown in Figure~\ref{subfig:dcaraw}, which can heavily influence the learning process of our models.
For this reason, our first step is to clean the data, as shown in Figure~\ref{subfig:dcaclean}.
To do so, we remove from the raw data the points in which the voltage is too close to the previous data point.
Indeed, as we can notice in Equation~\ref{eq:dca}, when the denominator is too close to zero, we will obtain values that are far too high and will constitute fake peaks in the plot.
Finally, to remove irregularities on the plot, we need to smoothen it as shown in Figure~\ref{subfig:dcasmooth}.
To do so, we use the Savitzky-Golay filter, which is a type of linear filter that uses a moving window of data to estimate the value of a given point in the signal~\cite{savitzky1964smoothing}.
These processing steps are based on the work of Thompson et~al.~\cite{thompson2020diffcapanalyzer} and are consolidated as best practices when dealing with discrete sampling of data associated with batteries~\cite{feng2020reliable, olson2023differential}.
After this procedure (which we apply to all the datasets), we ensure that our processed data do not contain any artifacts which might create biases during classification.

\begin{figure*}[!htpb]
  \centering
  \begin{subfigure}{0.33\textwidth}
     \centering
     \includegraphics[width=\textwidth]{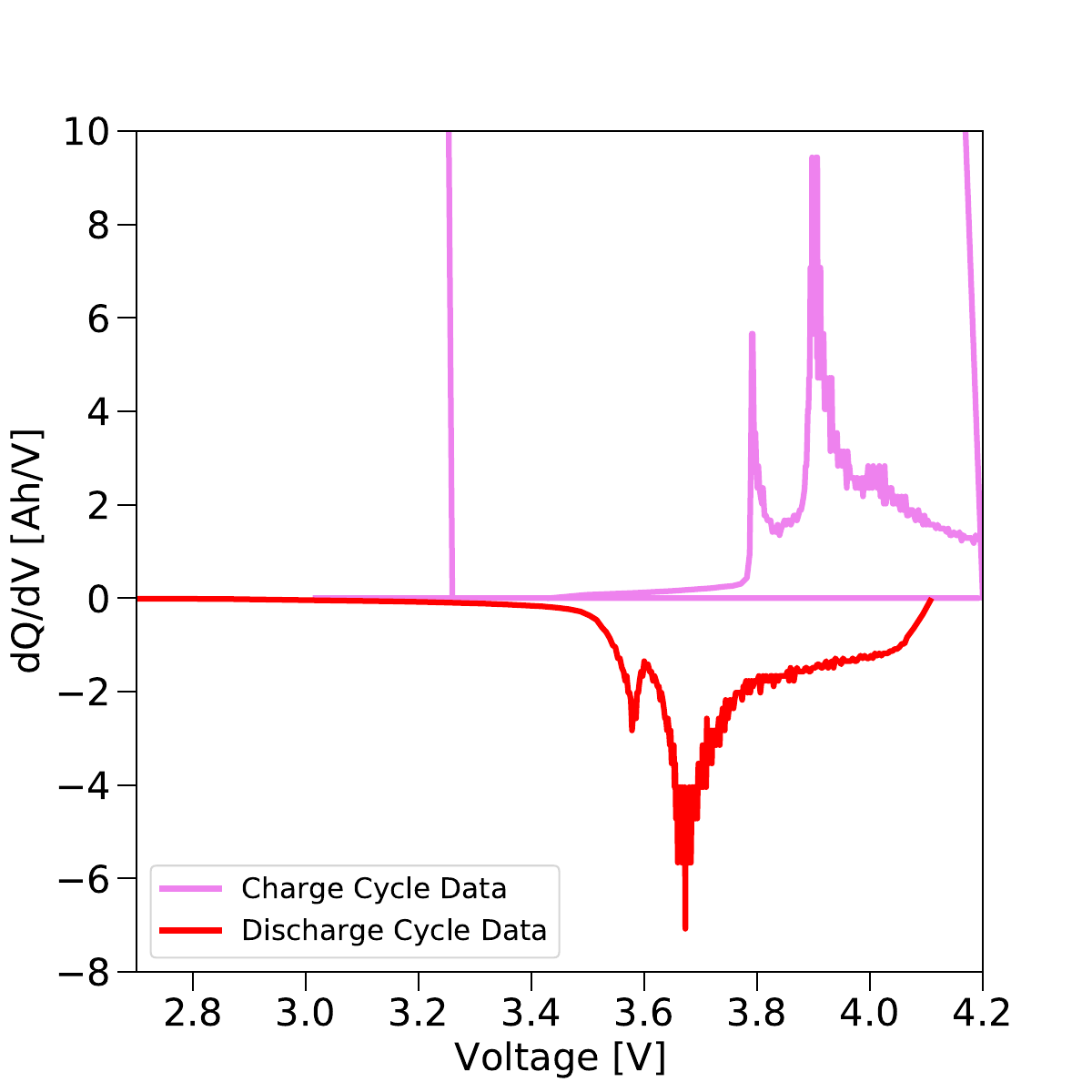}
     \caption{Raw DCA plot.}
     \label{subfig:dcaraw}
  \end{subfigure}
  \begin{subfigure}{0.33\textwidth}
     \centering
     \includegraphics[width=\textwidth]{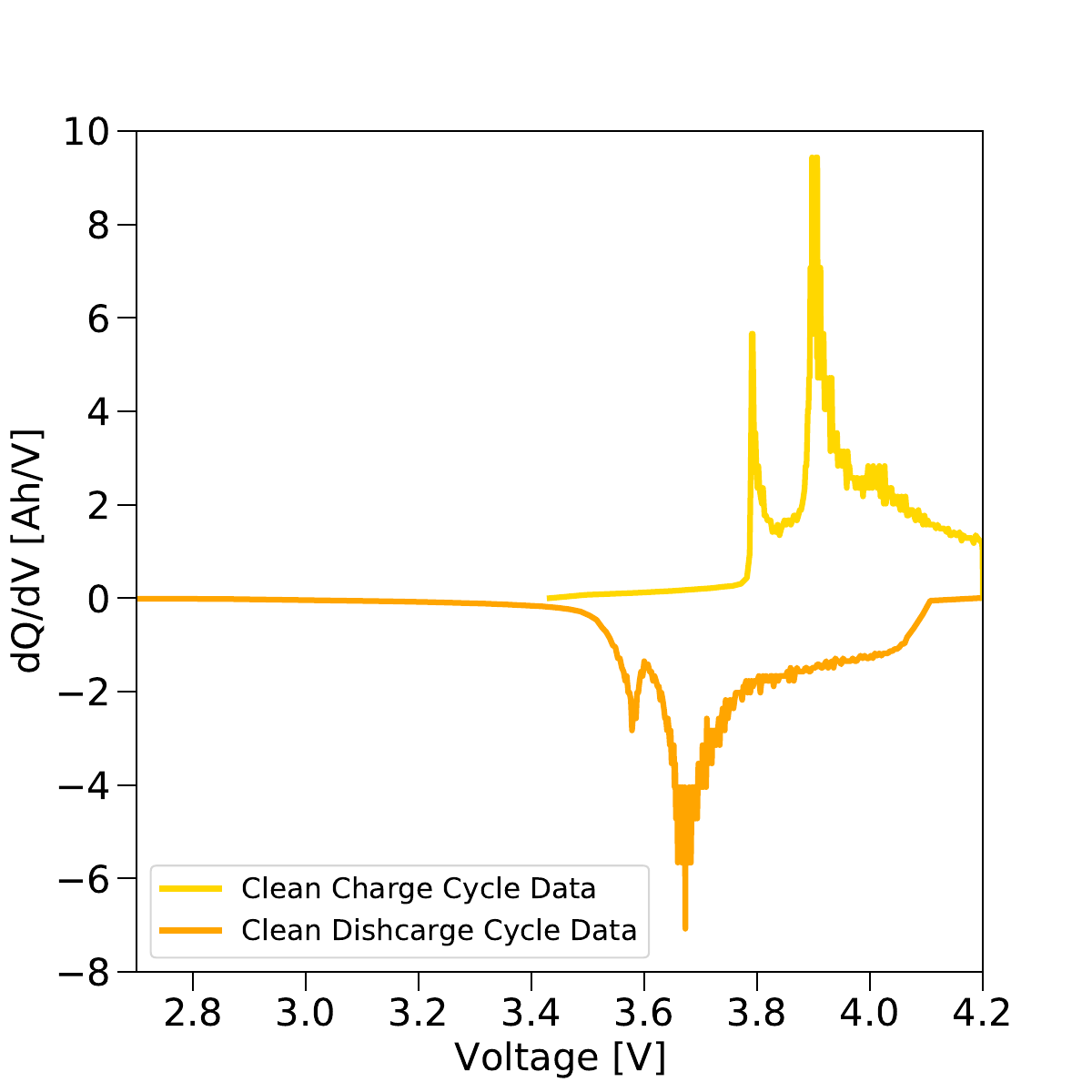}
     \caption{Clean DCA plot.}
     \label{subfig:dcaclean}
  \end{subfigure}
  \begin{subfigure}{0.33\textwidth}
     \centering
     \includegraphics[width=\textwidth]{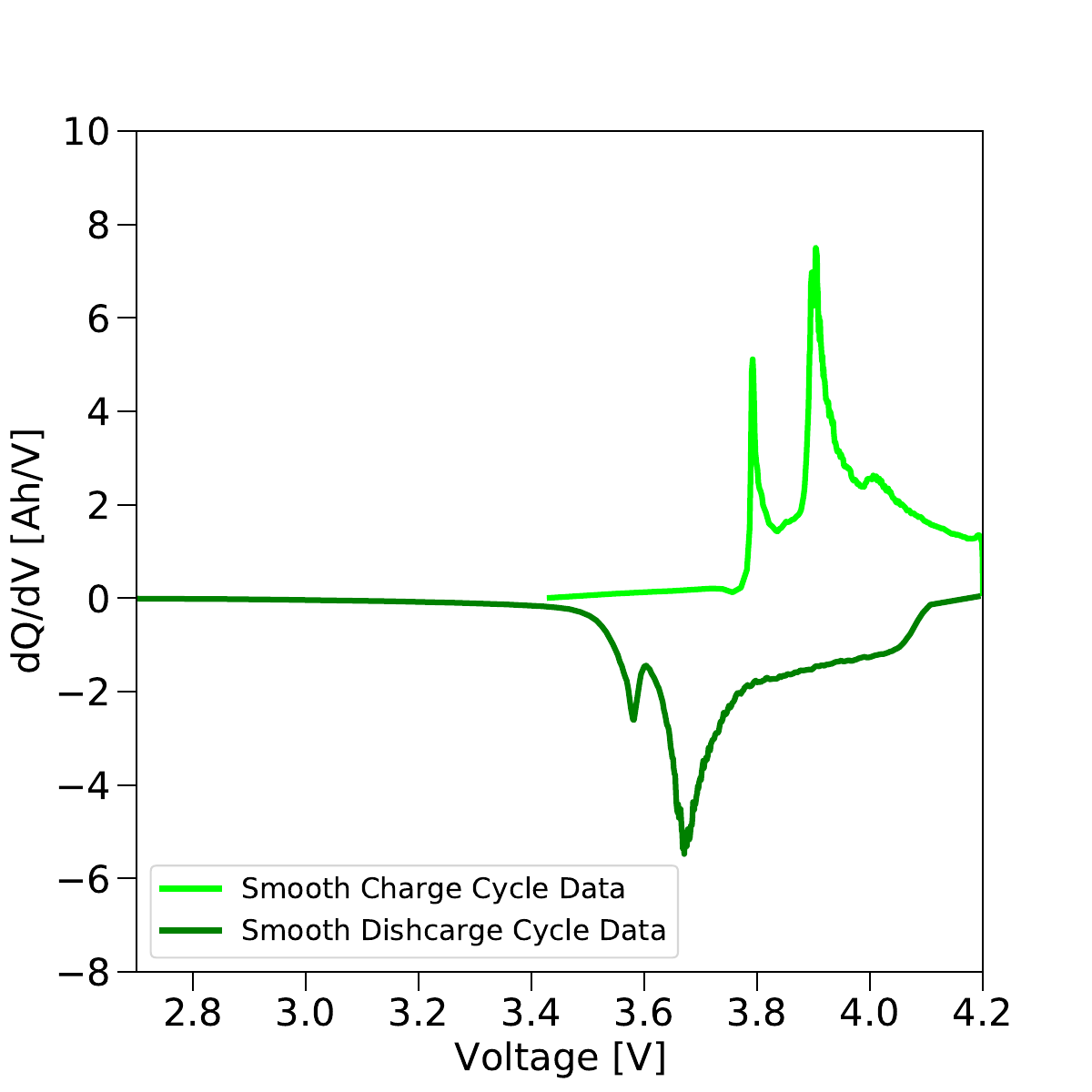}
     \caption{Smooth DCA plot.}
     \label{subfig:dcasmooth}
  \end{subfigure}
  \caption{Data processing overview on Differential Capacity Analysis.}
  \label{fig:dcaproc}
\end{figure*}
\section{Methodologies}
\label{sec:methodologies}

In this section, we outline in more detail the techniques that we use for authentication.
First, we discuss our feature extraction procedure, which is a crucial part of our work given the heterogeneity of the data we use (Section~\ref{subsec:feature}).
Then, we overview the Machine Learning models we use as classifiers and their optimal hyperparameter search (Section~\ref{subsec:models}). 

\subsection{Feature Extraction}
\label{subsec:feature}

In Section~\ref{subsec:processing}, we described our preprocessing procedure and how we extract and clean the data from our various datasets.
However, since we will use Machine Learning models to authenticate the various battery cells, graph data is not so suited in its raw form to be fed to the classifiers.
Moreover, in Section~\ref{subsec:dca} and~\ref{subsec:eis}, we have identified which could potentially be the most important characteristics in our data, i.e., peaks values and location for DCA and dependence on battery usage for EIS.
For these reasons, we define a common feature extraction procedure that will consider the many different aspects of the plotted data.
We can extract a constant number of features from any plot by using \textit{tsfresh}, an open-source Python package designed for time series feature extraction and selection~\cite{christ2018time}.
While our data is not time series data per se, tsfresh allows us to extract statistical features, spectral features, and features related to autocorrelation, trend, resampling, and others.

\subsection{Models}
\label{subsec:models}

Once the datasets have been processed and features have been extracted, we define eight different Machine Learning (ML) models to deploy for the authentication task.
Machine Learning is the field of Artificial Intelligence (AI) that focuses on developing algorithms that can learn from and make predictions based on data.
For this reason, it is often used in authentication~\cite{siddiqui2021user}.
Indeed, by identifying patterns in the DCA and EIS data, it is possible to identify not only the cell chemistry from which it has been extracted but also the specific samples among a pool of battery cells.
While DCA and EIS data has been widely used in ML models for SOC and SOH estimation, to the best of our knowledge, it has never been used for the authentication of the battery cells~\cite{kurzweil2022differential, babaeiyazdi2021state}.

In Table~\ref{tab:models}, we show the different models that we use.
We selected models commonly used in literature, particularly on tasks involving the same type of data that we use~\cite{zhu2019equivalent}.
Additionally, in order to optimize their performance, we perform Grid Search by defining possible values for the hyperparameters and training a model for each combination of values.
After cross-validating each model and determining its best combination of hyperparameters, we evaluate the best estimator on a test set.

The reader might notice that, in all our models, we did not include any deep or complex network.
Indeed, even the simple neural network that we consider, consists of only one hidden layer with a maximum size of 200.
While deep networks have been widely used in the literature with data concerning lithium-ion batteries~\cite{ma2022remaining, pepe2022neural}, we opted for more lightweight models both in terms of computational time and needed processing power.
Indeed, practical implementations of both DCAuth and EISthentication should be affordable and accessible on many different battery-powered devices, which might not possess the necessary hardware to run complex models, or might not be fast enough for authentication purposes.
Nonetheless, as will be shown in Section~\ref{sec:results}, the best models can still obtain high scores while maintaining their lightweight characteristics.

\def\modlen{4.15cm}
\def\parlen{3.6cm}
\def\onelinesmod{15pt}
\def\twolinesmod{25pt}
\def\threelinesmod{35pt}

\begin{table}[!htpb]
  \centering
  \caption{Machine Learning models deployed for classification and hyperparameters subject to Grid Search.}
  \label{tab:models}
  \begin{tabular}{p{\modlen}|p{\parlen}}
    \toprule
    \textbf{Models} & \textbf{Hyperparameters} \\
    \midrule
    
    \rowcolor{gray!15} AdaBoost (AB) & \parbox[][\onelinesmod][c]{\parlen}{
    $\bullet$ Number of estimators
    } \\\hline

    Decision Tree (DT) & \parbox[][\twolinesmod][c]{\parlen}{
    $\bullet$ Criterion\\
    $\bullet$ Maximum Depth
    } \\\hline

    \rowcolor{gray!15} Gaussian Naive Bayes (GNB) & \parbox[][\onelinesmod][c]{\parlen}{
    $\bullet$ Variance Smoothing
    } \\\hline

    Nearest Neighbors (KNN) & \parbox[][\twolinesmod][c]{\parlen}{
    $\bullet$ Number of neighbors\\
    $\bullet$ Weight function
    } \\\hline

    \rowcolor{gray!15} Neural Network (NN) & \parbox[][\threelinesmod][c]{\parlen}{
    $\bullet$ Hidden layer sizes\\
    $\bullet$ Activation function\\
    $\bullet$ Solver
    } \\\hline

    \parbox[][\twolinesmod][c]{\modlen}{Quadratic Discriminant Analysis (QDA)} & \parbox[][\twolinesmod][c]{\parlen}{
    $\bullet$ Regularization Parameter
    } \\\hline

    \rowcolor{gray!15} Random Forest (RF) & \parbox[][\twolinesmod][c]{\parlen}{
    $\bullet$ Criterion\\
    $\bullet$ Number of estimators
    } \\\hline

    Support Vector Machine (SVM) & \parbox[][\threelinesmod][c]{\parlen}{
    $\bullet$ Kernel\\
    $\bullet$ Regularization parameter\\
    $\bullet$ Kernel coefficient
    } \\
    
    \bottomrule
  \end{tabular}
\end{table}
\section{Evaluation}
\label{sec:results}

We now give an experimental evaluation of our methodologies on the proposed datasets.
First, we will disclose the metrics and testing scenarios (Section~\ref{subsec:metrics}).
Given the differences in data and processing, we differentiate the two methodologies and evaluate them separately in Section~\ref{subsec:dcauth} and~\ref{subsec:eisthentication}.

\subsection{Metrics}
\label{subsec:metrics}

We use four standard metrics to evaluate our models: accuracy, precision, recall, and F1 score.
By using True Positives (TP), True Negatives (TN), False Positives (FP), and False Negatives (FN), those metrics are defined as follows.
\begin{equation}
    Accuracy = \frac{TN + TP}{TP + TN + FP + FN},
\end{equation}
\begin{equation}
    Precision = \frac{TP}{TP + FP},
\end{equation}
\begin{equation}
    Recall = \frac{TP}{TP + FN},
\end{equation}
\begin{equation}
    F1\:\:Score = 2\frac{Precision \cdot Recall}{Precision + Recall}.
\end{equation}

While the main focus of our work is the authentication of battery models and architectures, we also study the ability of our models to perform identification.
Although the data and features for these two tasks are the same, some core conceptual differences set them apart.
\begin{itemize}
    \item \textbf{Authentication} -- 
    The process of verifying whether a battery is genuine or not.
    As a Machine Learning classification problem, this translates to a binary classification where the only two labels are \textit{authenticated} (or legitimate) and \textit{not authenticated}.
    \item \textbf{Identification} -- 
    The process of identifying the type and specifications of a battery.
    As a Machine Learning classification problem, this translates to a multiclass classification where each label constitutes a battery model or architecture.
\end{itemize}
Since our datasets are composed of data retrieved from original battery samples cycled in their intended application, in the provided evaluation, a counterfeit model is equivalent to a different battery model.
When dealing with authentication, in particular, we are interested in two more metrics: False Acceptance Rate (FAR) and False Rejection Rate (FRR).
FAR is the rate at which the authentication system incorrectly accepts an invalid or unauthorized battery as a valid battery.
FRR is the rate at which the authentication system incorrectly rejects a valid battery as an invalid battery.
These metrics are defined as follows.
\begin{equation}
    FAR = \frac{FP}{FP + TN},
\end{equation}
\begin{equation}
    FRR = \frac{FN}{FN + TP}.
\end{equation}

\subsection{DCAuth}
\label{subsec:dcauth}

We start our evaluation by first measuring the F1 score of all our ML models in the identification task for DCAuth.
To do that, we split the whole dataset into a training set and a test set with respective percentages of 80\% and 20\%.
Further splits to include a validation set are not needed since we are performing 5-fold cross-validation during our Grid Search.
When dealing with identification, we perform multiclass classification where the samples for each label have been collected from various datasets.
This leads to an imbalanced distribution of the labels, which can affect our processing and generate bias in our results.
For this reason, we first balance the number of labels in the dataset by performing random undersampling~\cite{lemaitre2017imbalanced}.
While oversampling techniques, such as SMOTE~\cite{chawla2002smote}, might help us in pursuing the same objective without removing samples from our dataset, we found out that it negatively affected the performance of the classifier, probably due to the high number of features.
Results for both architecture authentication and battery model authentication are shown in Figure~\ref{fig:dcauth-ident}.
While most ML models manage to achieve good results, the Random Forest classifier appears to be the best one, obtaining an F1 score close to 1 in battery architecture identification (0.99) and 0.92 in battery model identification.
With the sole exception of the Gaussian Naive Bayes classifier, all ML models obtain better results in battery architecture identification than battery model identification.
Indeed, the number of architectures is far less than that of battery models (respectively, 5 against 11).
Furthermore, battery cells that share the same architecture usually have similar specifications.
Thus, DCA plots between cells are comparable, and the extracted features appear to be correlated with each other.
More details on the other metrics of evaluation can be found in our repository.

\begin{figure}[!htpb]
    \centering
    \includegraphics[width=\linewidth]{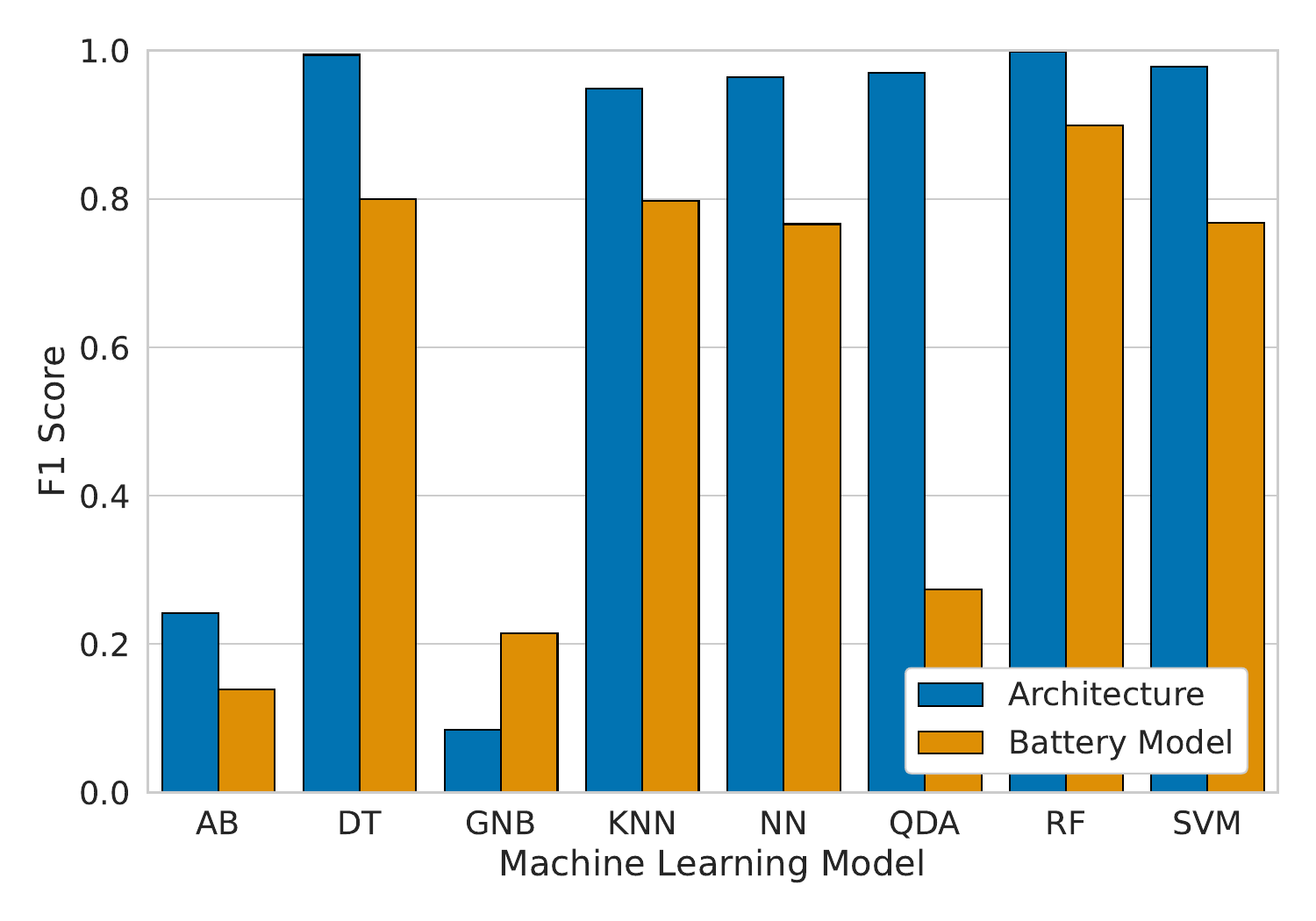}
    \caption{Identification of architectures and battery models in DCAuth.}
    \label{fig:dcauth-ident}
\end{figure}

In the previous evaluation, we considered the dataset composed of data retrieved from both charging and discharging cycles.
While also these two types of cycles are equally distributed, their individual effect on the ML model performance might be different.
To study this, we divide the dataset into two different parts, one considering charging cycles and the other one considering only discharging cycles.
After that, we train different classifiers on each of them and compare the results, which are shown in Figure~\ref{fig:dcauth-ident-sep}.
As we can see, features extracted from the discharging cycles tend to perform slightly better in most of the models (Random Forest included).
Indeed, as seen in Table~\ref{tab:dca}, while charging protocols are often similar in different studies (e.g., CCCV charge), discharging techniques instead are more varied and might even depend on dynamic profiles (i.e., driving profiles for EV batteries).
Nonetheless, results are still comparable to the ones obtained when considering the whole dataset.

\begin{figure}[!htpb]
    \centering
    \includegraphics[width=\linewidth]{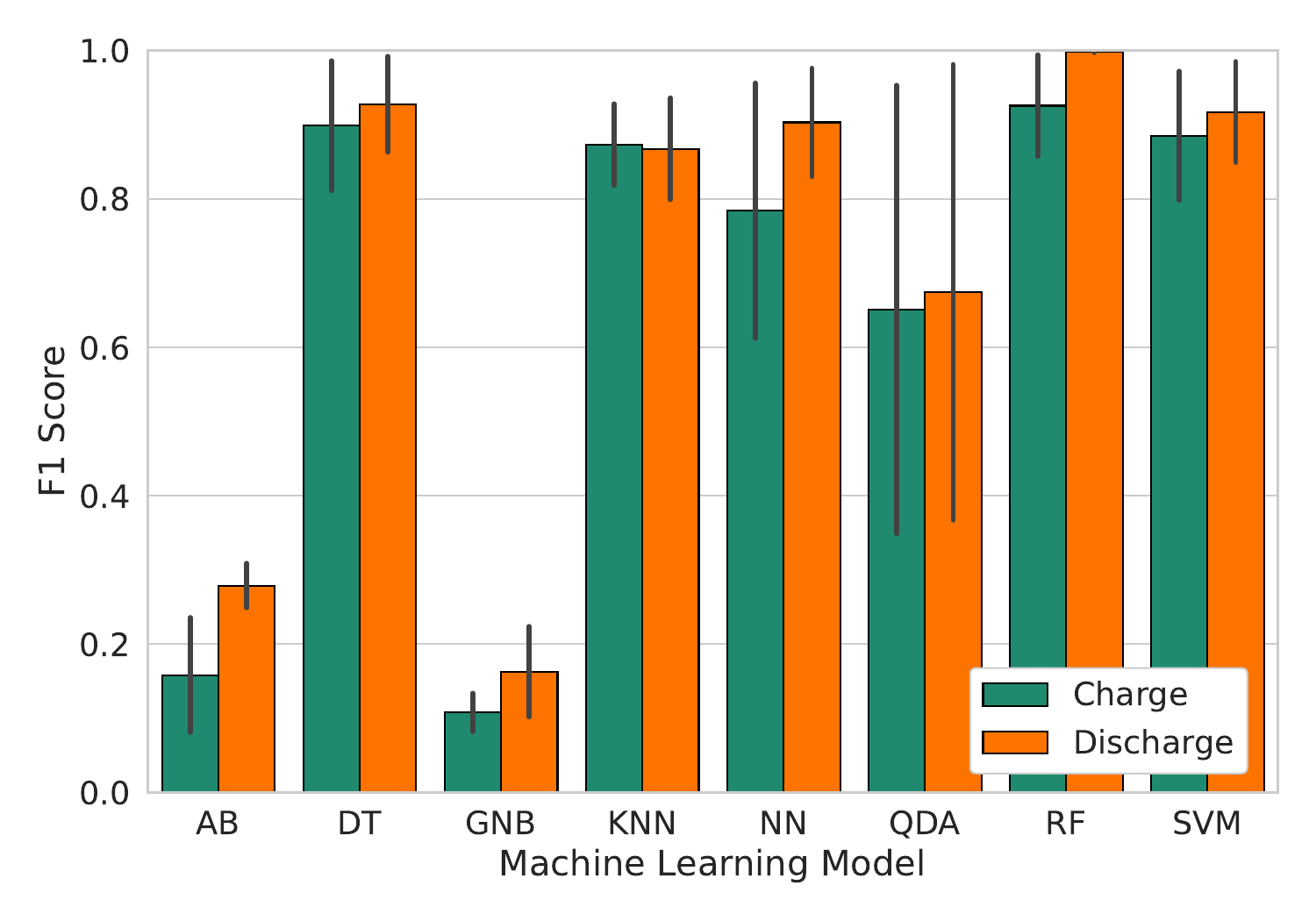}
    \caption{Identification in DCAuth when considering separate charging and discharging cycles. Results are averaged with respect to architecture identification and battery model identification.}
    \label{fig:dcauth-ident-sep}
\end{figure}

We now move to the authentication task, which effectively translates to binary classification.
However, given the number of labels in the dataset, we must consider the effect that an imbalanced distribution can have on the evaluation results.
Indeed, despite the undersampling performed in the identification task, when considering just one label as legitimate and all the others as the not authorized ones, we end up with heavy imbalances in the data distribution.
This is expected since many security and privacy scenarios are inherently imbalanced.
Two examples are hate speech detection, where datasets are often imbalanced towards the hateful class~\cite{grondahl2018all}, and intrusion detection systems, where rarer attacks often constitute the minority classes~\cite{bedi2021siamids}.
Thus, we define four different levels of imbalances with respect to legitimate and not authorized classes.
\begin{itemize}
    \item \textbf{50/50} -- Samples equally split between the two labels.
    \item \textbf{40/60} -- 40\% legitimate, 60\% not authorized.
    \item \textbf{30/70} -- 30\% legitimate, 70\% not authorized.
    \item \textbf{20/80} -- 20\% legitimate, 80\% not authorized.
\end{itemize}
As a first type of evaluation for the authentication task, we train all our ML models on all our dataset balances (obtaining a total of $8\cdot4=32$ models for each task) and compute their F1 score on the test set.
While other metrics are published in our repository, we focus on the F1 score in particular since it combines both precision and recall, and we are dealing with imbalanced distribution.
Results averaged with respect to the balance levels are shown in Figure~\ref{fig:dcauth-auth}.
These results are acquired by averaging F1 scores from training and testing each model with diverse combinations of genuine and fake battery labels, where a label is treated as legitimate while others as counterfeit.
By comparing these results with the ones shown in Figure~\ref{fig:dcauth-ident}, we notice similar values with the exception of some of the ML models, which are able to obtain better results in a binary classification scenario.
Nonetheless, the Random Forest classifier is still the best among the considered ones.

\begin{figure}[!htpb]
    \centering
    \includegraphics[width=\linewidth]{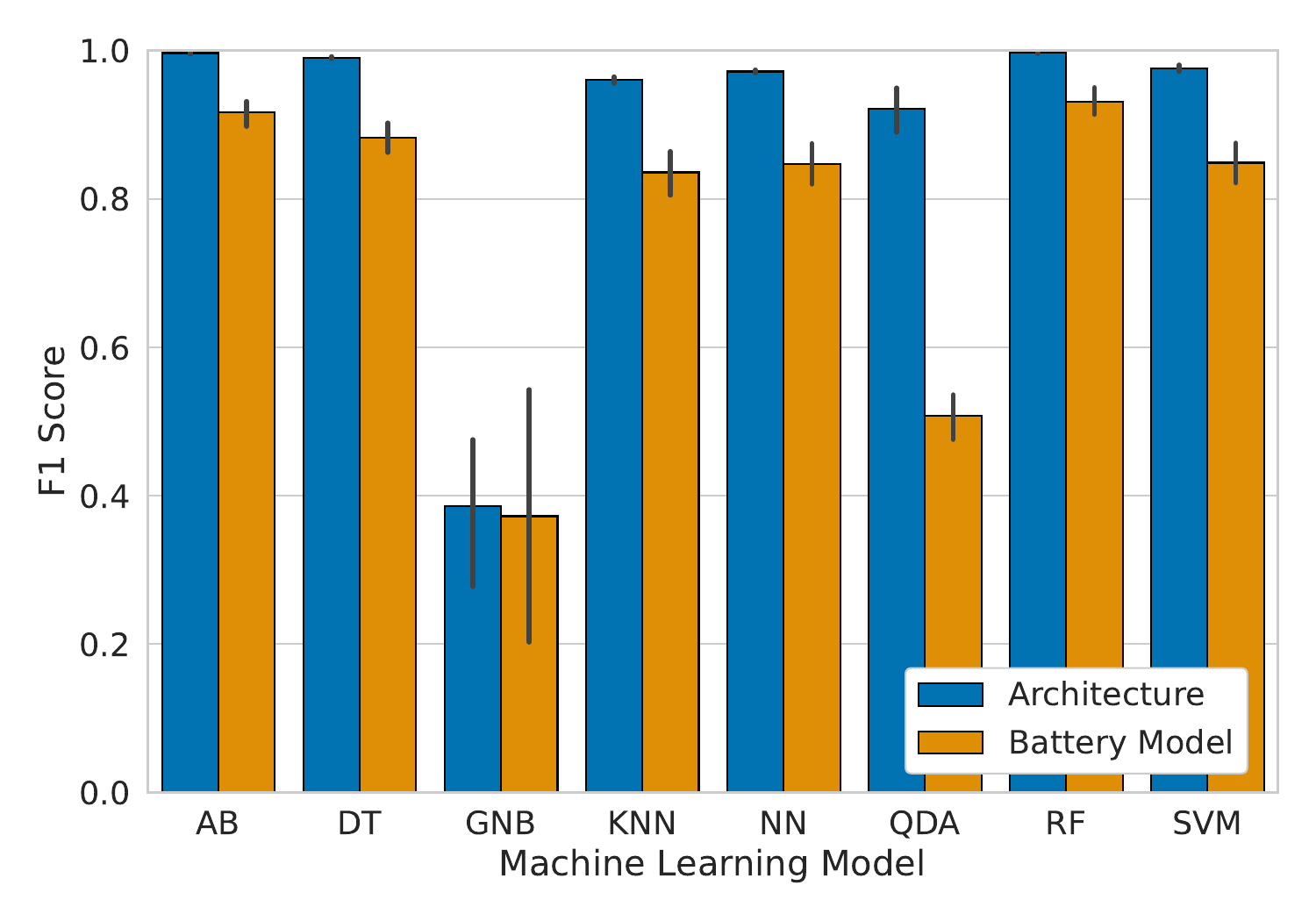}
    \caption{Authentication of architectures and battery models in DCAuth. Results are averaged with respect to the different balance levels for the dataset distribution.}
    \label{fig:dcauth-auth}
\end{figure}

As a second type of evaluation, we study the effect of the different dataset imbalances on the False Acceptance Rate and False Rejection Rate of our models.
To do so, we need to average results for architecture authentication and battery model authentication.
This is a fair assumption since, for almost all models, the performances of the two tasks are comparable.
Results for both FAR and FRR are shown in Figure~\ref{fig:dcauth-auth-farfrr}.
Unsurprisingly, models performing poorly in authentication have higher FARs and FRRs (e.g., GNB and QDA).
It is interesting instead to analyze the trend of those rates across the different distributions.
Indeed, while good performers such as RF and AB seem to be almost unaffected by the distribution given their excellent baseline performance, other models seem to present higher or lower performances depending on the balance level and term of evaluation.
In particular, FAR is generally higher whenever the distribution is more unbalanced towards the negative class, while FRR has an opposite tendency.
This is due to the fact that False Positives are usually more frequent whenever the negative class is the predominant one, and vice versa for False Negatives.  
\begin{figure*}[!htpb]
  \centering
  \begin{subfigure}{0.498\textwidth}
     \centering
     \includegraphics[width=\textwidth]{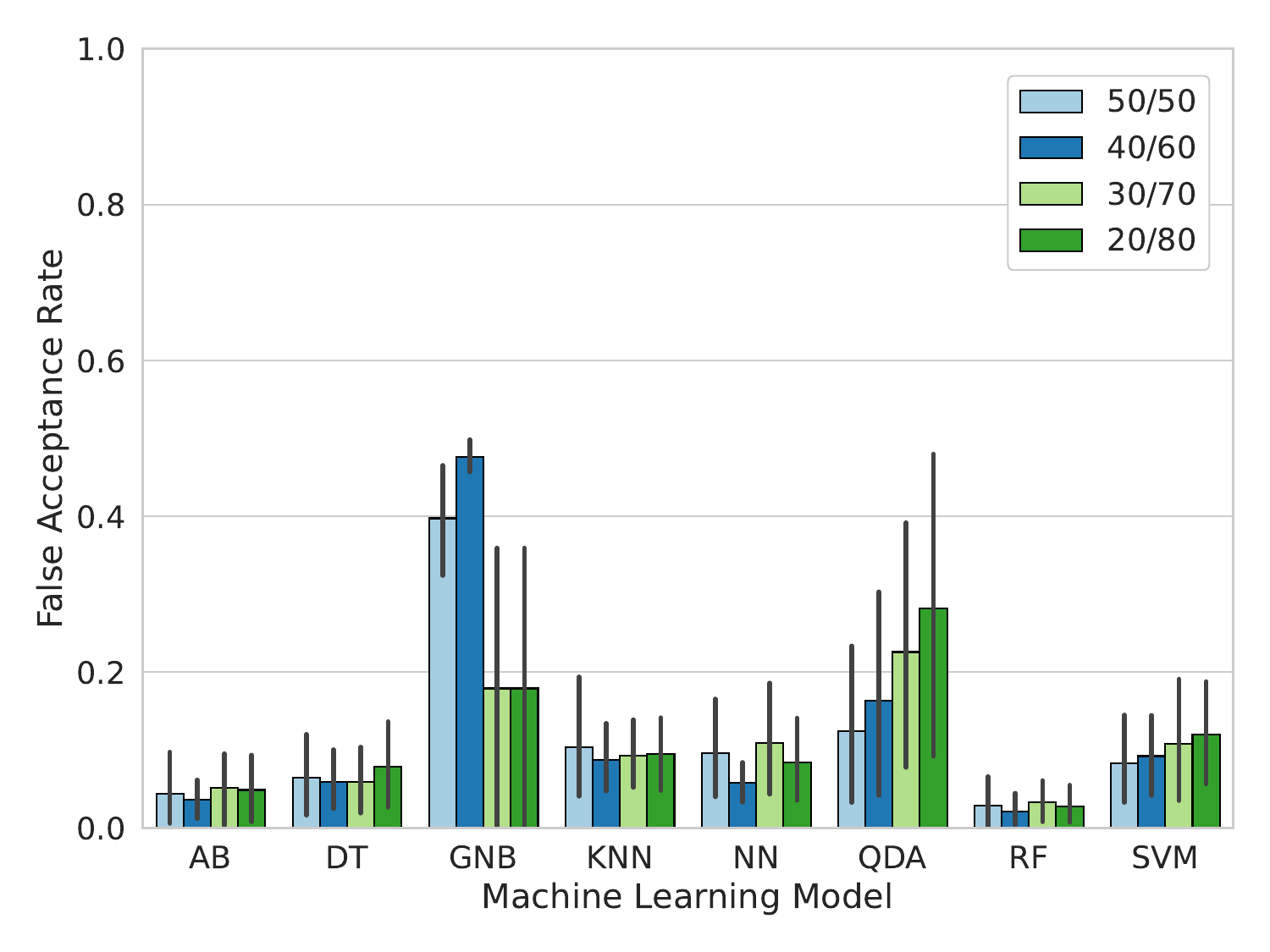}
     \caption{False Acceptance Rate.}
     \label{subfig:dcauth-far}
  \end{subfigure}
  \begin{subfigure}{0.498\textwidth}
     \centering
     \includegraphics[width=\textwidth]{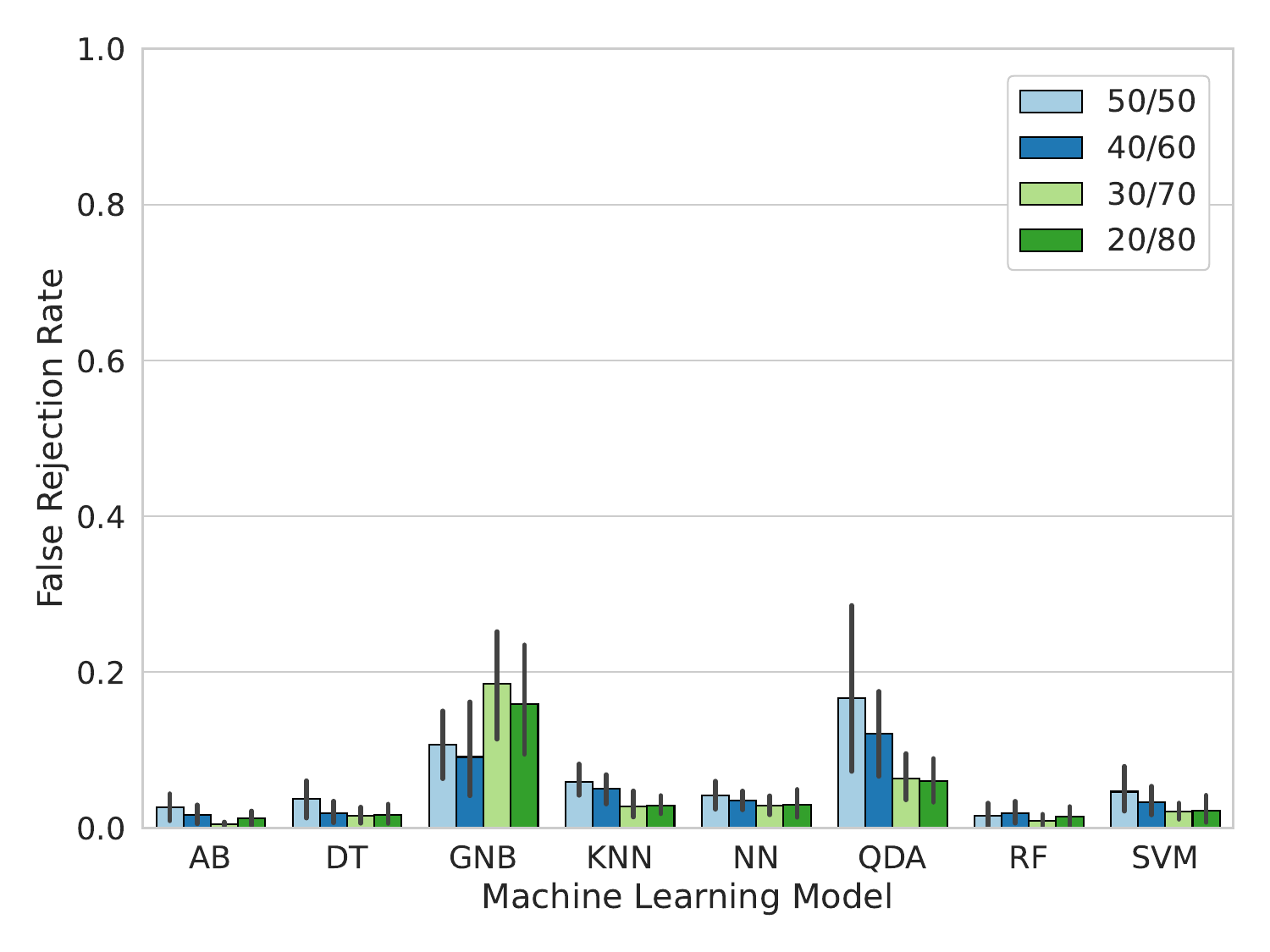}
     \caption{False Rejection Rate.}
     \label{subfig:dcauth-frr}
  \end{subfigure}
  \caption{Authentication in DCAuth for each dataset distribution. Results are averaged with respect to architecture authentication and battery model authentication.}
  \label{fig:dcauth-auth-farfrr}
\end{figure*}
Nonetheless, using a balanced dataset seems to be the optimal solution for two main reasons.
\begin{enumerate}
    \item The effect of the dataset distribution is greater while evaluating FAR with respect to FRR.
    This means that while the False Rejection Rate might slightly increase with respect to more imbalanced distributions, an imbalanced dataset would otherwise greatly (and negatively) impact the overall performance of the model.
    \item The aim of the task is to authenticate one (or more) battery cell in a system.
    Thus, in order to avoid putting the users at risk, False Acceptance events are more critical than False Rejection events.
    Nevertheless, having a good variety of samples in the not authorized label can greatly improve the learning process of the models.
\end{enumerate}

A similar analysis to the one shown in Figure~\ref{fig:dcauth-ident-sep} (i.e., splitting the datasets in charging and discharging cycles) can also be performed for the authentication task.
Given the similarity of the obtained scores, we refrain from including an additional plot to show their values.

\subsection{EISthentication}
\label{subsec:eisthentication}

We now move to the evaluation of EISthentication.
While the types of evaluations are similar to the ones performed for DCAuth, the characterization of the data is slightly different.
Indeed, before splitting the dataset into training set and test set (with the same percentages of DCAuth and using the same 5-fold cross-validation during Grid Search) we filter the features with the \textit{tsfresh} package.
Its \texttt{feature\_selection} package includes a selection method that evaluates the importance of the different extracted features, and thus we keep only the most relevant ones in our datasets.
In DCAuth, this procedure does not affect the size of the dataset so much (from 788 features to around 750).
In EISthentication, instead, the number of filtered features is almost half that of original features (from 788 to around 400).
This could indicate that EIS data might contain some redundancy, and thus results might differ.

As with DCAuth, we start by evaluating the baseline performance of the identification task on both architecture identification and battery model identification.
Results are shown in Figure~\ref{fig:eis-ident}.
Some differences can be noted while comparing these results with the ones obtained with DCAuth in Figure~\ref{fig:dcauth-ident}.
First, different models produce more consistent results, while in DCAuth, some of them (AB and GNB in particular) performed particularly poorly in both tasks.
Secondly, there seems to be a bigger gap in the F1 score between architecture identification and battery model identification.
This could mean that the characteristics related to the physical structure and chemistry of the battery extracted through EIS are more directly related to its architecture than its modeling.
Lastly, the Random Forest classifier is still the best one among the models considered and reaches F1 scores (0.96 for architecture identification and 0.88 for battery model identification) close to the ones obtained in DCAuth.

\begin{figure}[!htpb]
    \centering
    \includegraphics[width=\linewidth]{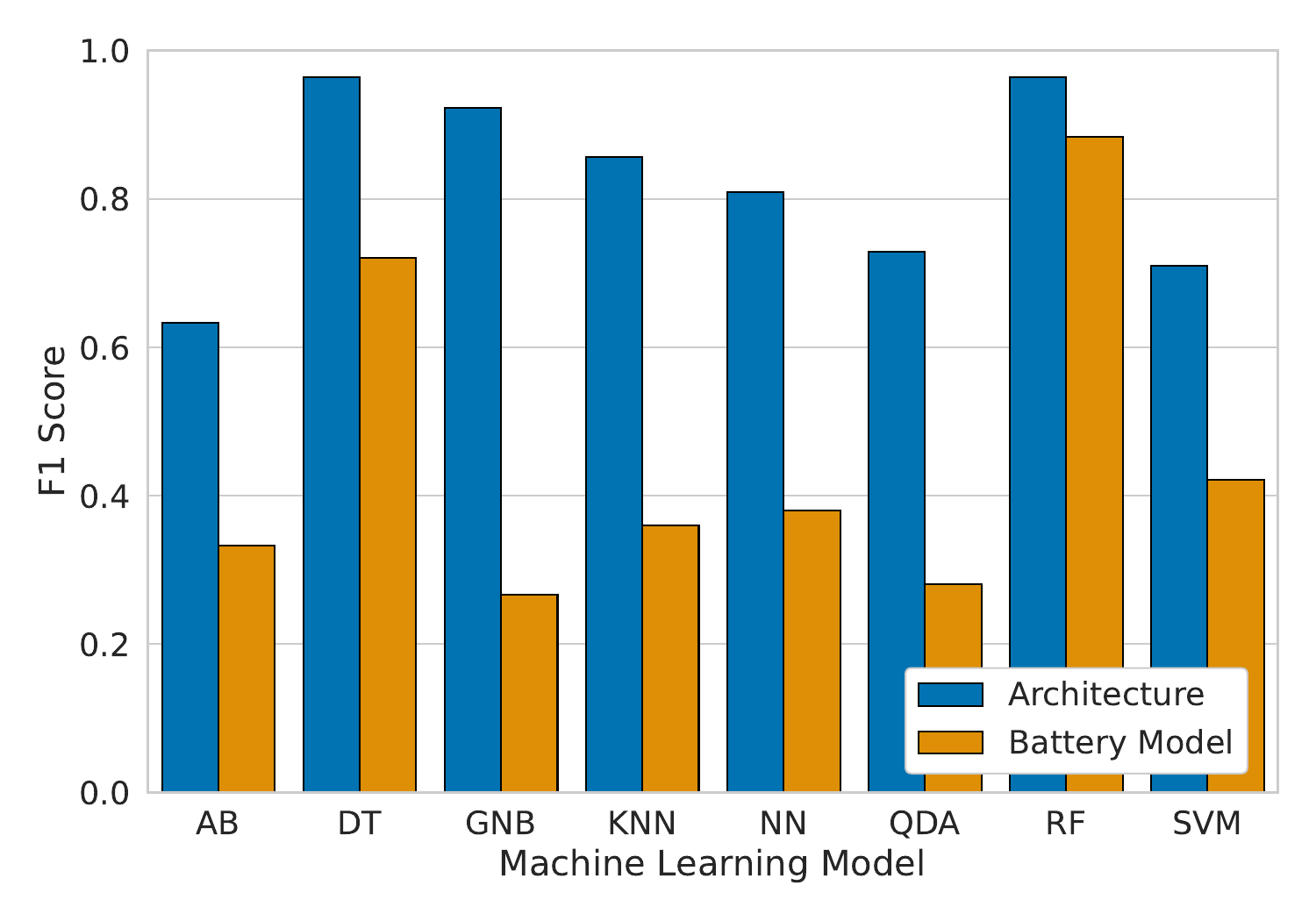}
    \caption{Identification of architectures and battery models in EISthentication.}
    \label{fig:eis-ident}
\end{figure}

Moving to authentication, we notice similar behaviors in the results, which are shown in Figure~\ref{fig:eis-auth}.
As with DCAuth, some of the models that had poor performance in identification are now obtaining higher results.
Instead, other models that were already performing well in identification now reach approximately the same F1 scores.

\begin{figure}[!htpb]
    \centering
    \includegraphics[width=\linewidth]{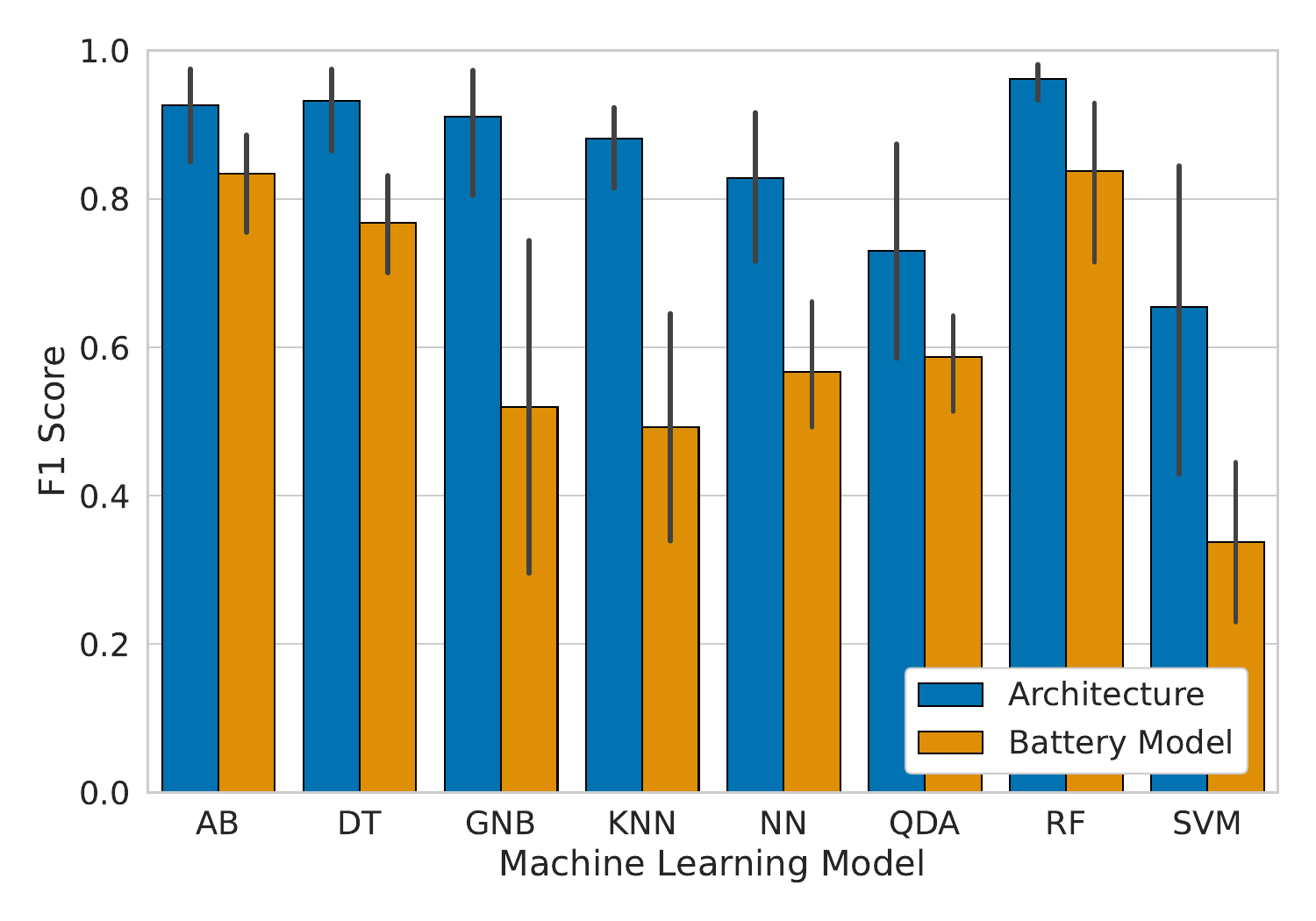}
    \caption{Authentication of architectures and battery models in EISthentication. Results are averaged with respect to the different balance levels for the dataset distribution.}
    \label{fig:eis-auth}
\end{figure}

Even with EISthentication, in the authentication task we are using the same four balance levels for the dataset distribution.
Therefore, we can analyze the trend of the False Acceptance Rate and False Rejection Rate at the varying of those balance levels.
Results are shown in Figure~\ref{fig:eis-auth-farfrr}.
We can notice the same trends that were present with DCAuth in Figure~\ref{fig:dcauth-auth-farfrr}, but they appear to be more uniform across the different models.
However, more imbalanced distributions appear to affect both the FAR and the FRR of the models more with respect to DCAuth.
Thus, all the more so, in this case, the optimal distribution would be the balanced one for the reasons discussed above.
\begin{figure*}[!htpb]
  \centering
  \begin{subfigure}{0.498\textwidth}
     \centering
     \includegraphics[width=\textwidth]{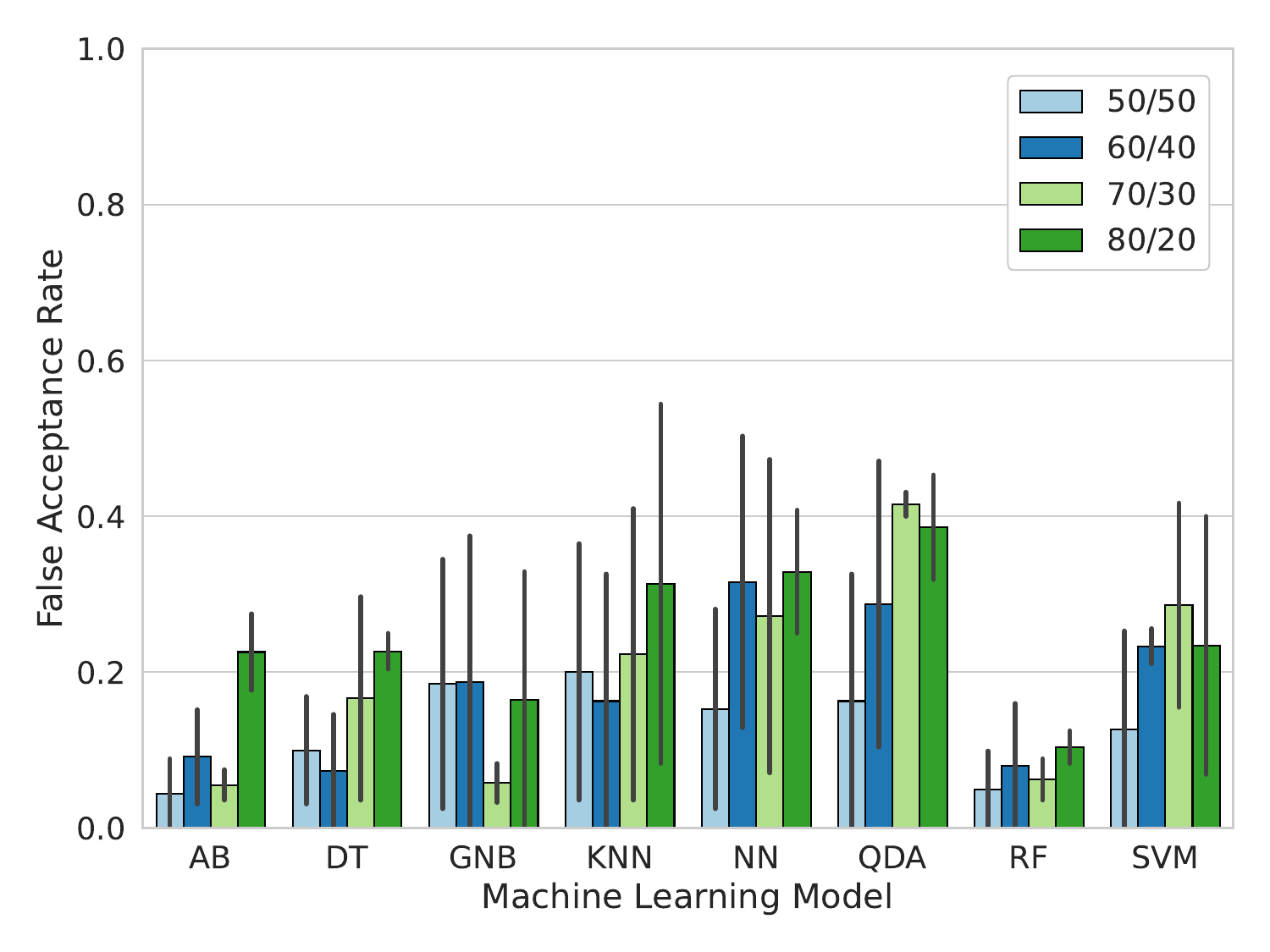}
     \caption{False Acceptance Rate.}
     \label{subfig:eis-far}
  \end{subfigure}
  \begin{subfigure}{0.498\textwidth}
     \centering
     \includegraphics[width=\textwidth]{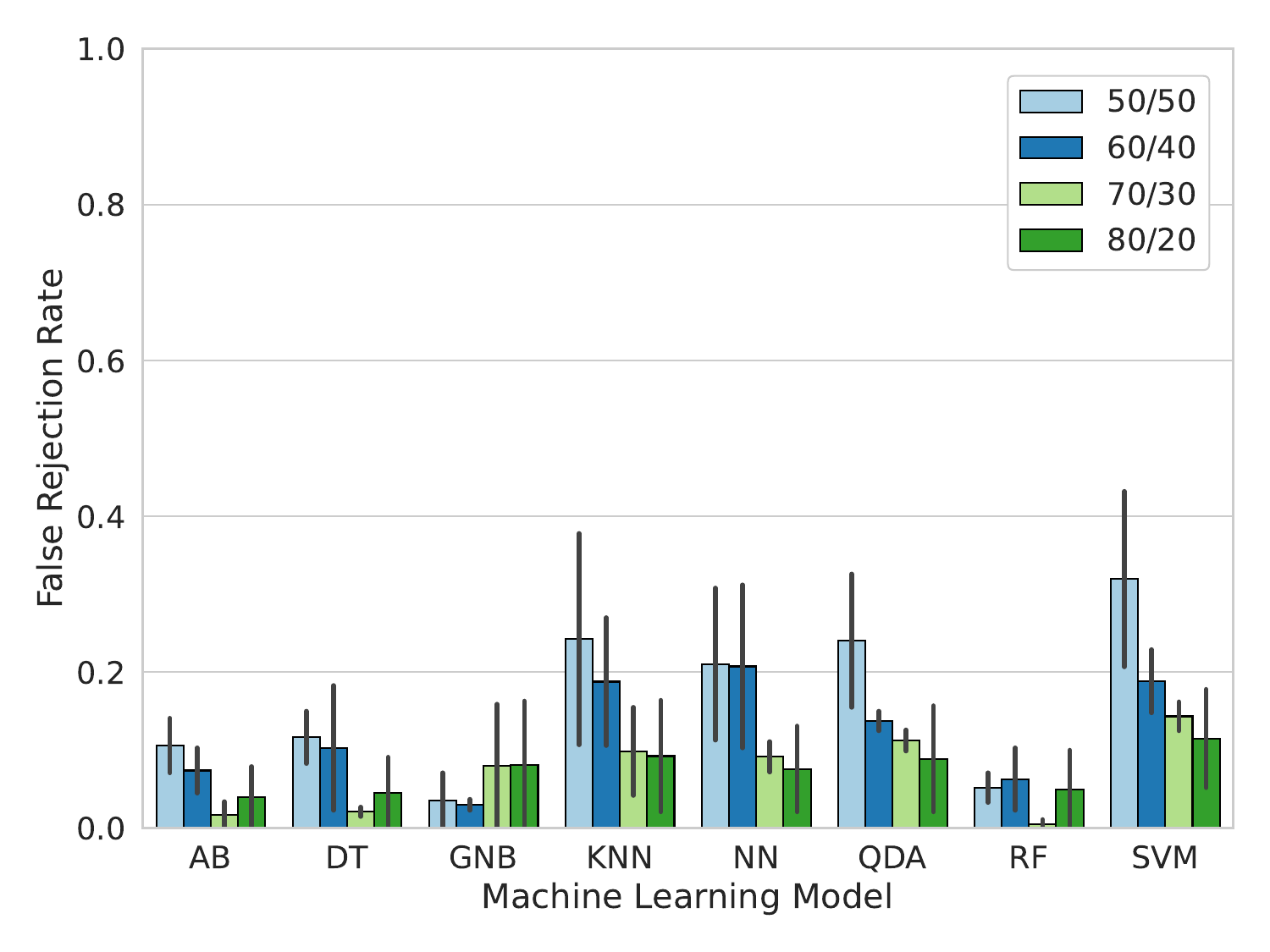}
     \caption{False Rejection Rate.}
     \label{subfig:eis-frr}
  \end{subfigure}
  \caption{Authentication in EISthentication for each dataset distribution. Results are averaged with respect to architecture authentication and battery model authentication.}
  \label{fig:eis-auth-farfrr}
\end{figure*}
\section{Discussion}
\label{sec:discussion}

In this section, we will analyze the obtained results and discuss the possible criticalities of our methodologies.
We first analyze the models' performances and use explainable machine learning techniques to have a more detailed insight into the importance of the features (Section~\ref{subsec:featureimportance}).
A comparison of the two presented methodologies is detailed in Section~\ref{subsec:comparison}, while potential limitations are highlighted in Section~\ref{subsec:limitations}.
Finally, we discuss the possibility of evasion attacks on the Machine Learning models in Section~\ref{subsec:evasion}.

\subsection{Feature Importance}
\label{subsec:featureimportance}

In Section~\ref{subsec:dca} and Section~\ref{subsec:eis}, we detailed the techniques used to retrieve data from the battery cells and made some hypotheses on their goodness for classification.
In particular, we speculate that the moving peaks in the DCA plots might be used for authentication and for EIS instead the relationship between different environmental conditions.
To confirm our hypothesis, we use Explainable Machine Learning (XAI) techniques on the ML models that we previously trained~\cite{roscher2020explainable}.
We use different techniques on the Random Forest classifier since it is the best performing model in all our tests.
Specifically, we use two techniques: Mean Decrease in Impurity (MDI) and SHapley Additive exPlanations (SHAP).
The former is a model-specific, global technique that computes the average decrease in Gini impurity or entropy that results from splitting a node using the specific features~\cite{han2016variable}.
Being model-specific, this technique should give us an accurate insight into the model and be faster to compute.
On the other hand, this measure suffers from so-called feature selection bias, i.e., it may erroneously assign high MDI values to features that are not highly correlated to the output.
For this reason, we also use SHAP, which on the contrary, is model-agnostic~\cite{lundberg2017unified}.

For DCAuth, we obtained similar results with both XAI techniques, which means that the 20 most important techniques extracted with each method greatly overlap.
The most important features appear to be related to Continuous Wavelet Transform (CWT) coefficients.
It is a technique used in signal processing to analyze the frequency content of a signal and can be used to identify patterns and features in the signal.
Another one of the most important features is the number of peaks in the time series, followed by the count of observed values in a specified interval.
Lastly, there is also autocorrelation, which is somehow expected since we are dealing with DCA plots in which different behaviors at the early stages of charging/discharging might already indicate a different battery cell or architecture.
Combining these features might show that peaks have an important role in the classification and thus might confirm our hypothesis.
Moreover, the difference between the average impacts that the features have on the model is small enough to consider them equally important ($\sim0.01$ in accuracy). 

Similar results are obtained with EISthentication.
While autocorrelation and the number of peaks were still present in the top 20 most important features, one of the most important ones appears to be the quantile feature.
The feature is often present multiple times with different $q$ values in all our models.
This could mean that the distribution of the data points is skewed (we performed the analysis in the balanced scenario) and certain percentiles are more meaningful than others.
We also find many Continuous Wavelet Transform (CWT) coefficients and Fast Fourier Transform (FFT) coefficients.
Thus, it appears that the model is more interested in the particular distribution of the data points than its specific peak values, indicating that variations in the different data points are more homogeneous on the whole space.

\subsection{Methodology Comparison}
\label{subsec:comparison}

In Section~\ref{subsec:dcauth} and Section~\ref{subsec:eisthentication}, we show that both our methodologies are able to obtain high scores in both identification and authentication.
What sets these methodologies apart is, of course, the type of data that they use for the processing and, thus, their possible scope of application.
Indeed, while DCAuth is the technique that manages to obtain the best results, to perform a prediction, it needs a full charge or discharge cycle for the battery.
While this assumption can be fair in some implementations, it has the drawback of being time-consuming (depending on the battery capacity) and degrading the State Of Health of the battery.
Indeed, each cycle affects the overall Remaining Useful Life (RUL) of a battery.
Nevertheless, some batteries are designed to be stressed for such a large number of cycles that a complete charge or discharge has little to no effect on its SOH.
Instead, if the capacity of the battery is too big to wait for a full cycle to be completed, EIS is generally faster.
The duration of EIS testing for a lithium-ion battery can vary depending on the chosen frequency range and required accuracy level.
For example, a typical EIS measurement for a lithium-ion battery might take around 10 to 30 minutes to complete.
Moreover, optimal results for EIS are produced when the battery has been at rest for long periods of time, which incentivize its authentication before its initial usage~\cite{leduc2020real}.
While performances in authentication might be slightly lower than DCAuth, EISthentication can be a valuable technique given the increasing availability of EIS measurements or estimations in different systems \cite{troltzsch2006characterizing, howey2013online, lohmann2015employing}.
Both these methodologies have also proven to be computationally lightweight, making their usage affordable on many different low-end devices.

\subsection{Limitations}
\label{subsec:limitations}

In our work, we considered a total of 20 datasets, accounting for 17 different battery models and 5 different battery architectures.
Instead, in real-world applications, millions of batteries might potentially be used for a specific scenario, and thus, the number of possible counterfeit samples becomes even larger.
In practice, this could increase the number of false positives and false negatives in our approaches, which could affect their accuracy, leading to some issues if applied to critical contexts.
To partially address this, selecting a subset of authorized battery models might be possible and then appropriately adjusting the dataset balance by incorporating data labeled as "counterfeit" from different battery models.
Moreover, it is unnecessary to consider every battery model for authentication in a single application.
For example, though a 18350 battery can be disguised as a 18650 battery through rewrapping, the opposite scenario is physically impossible.
Furthermore, adopting such a strategy to conceal a counterfeit battery wouldn't yield economic advantages for an attacker seeking to market an inferior battery at the cost of a genuine one.

Another aspect that could potentially alter performance with respect to the results shown in this paper is the aging of the battery cells.
In our work, we collected datasets that contain data retrieved from batteries cycled in many different stages of their lifespan, as described in Table~{\ref{tab:dca}} and Section~{\ref{subsub:eisdata}}.
However, as also shown in Figure~{\ref{fig:eis}}, several aspects such as SOH and SOC greatly affect the data distribution from which we extract features.
Thus, data extracted from batteries cycled in a limited set of conditions might not be authenticated if the dataset used to train the models did not account for that specific combination of parameters.
To account for this, we advise the usage of varied datasets for training and collecting data from batteries in many stages of their lifecycle.

\subsection{Evasion Attacks}
\label{subsec:evasion}

By only leveraging the electrochemical properties of the battery, DCAuth and EISthentication are resilient against most attacks while also providing scalability.
However, to achieve these results, our methodologies use Machine Learning models, which have been proven vulnerable to adversarial attacks~\cite{goodfellow2014explaining}.
In particular, evasion attacks represent a subset of adversarial attacks focused on altering input data to mislead Machine Learning models.
These attacks could potentially affect the authentication results of our methodologies since even our best classifier (i.e., Random Forest) has been shown to be vulnerable under specific conditions~\cite{apruzzese2018evading, zhang2020efficient}.
However, considering our system model, the implementation of these attacks in most applications is not currently practical due to the following reasons.
\begin{itemize}
    \item \textbf{Attacker Knowledge} -- To successfully evade authentication, the attacker needs access to the inner parameters of the target Machine Learning model.
    Since this information is usually hidden from the attacker, real-world attacks could differ significantly from the scenarios outlined in the literature~\cite{apruzzese2022real}.
    \item \textbf{Model Obfuscation} -- To gain access to the model or its training dataset, attackers can use techniques such as model inversion.
    However, procedures such as model obfuscation can prevent it and protect the model parameters from unauthorized access.
    \item \textbf{Adversarial Transferability} -- The knowledge that an attacker can gain to evade detection is not limited to the model architecture.
    By leveraging the dataset distribution and ground truth balance, malicious actors can craft evasion attacks that are "transferable" among different models.
    However, it has been shown how surrogate models struggle to evade most models impactfully~\cite{alecci2023your}.
\end{itemize}
\section{Conclusions}
\label{sec:conclusions}

In conclusion, the rise of counterfeit Li-ion batteries is a serious issue that demands the attention of the battery industry.
Counterfeit batteries can cause significant safety hazards and economic losses, and detecting them is becoming increasingly difficult as counterfeiters become more sophisticated.
Furthermore, current techniques for battery authentication can be fooled by various attacks or are not scalable to several battery models and architectures.

\paragraph{Contribution}
In this paper, we explored the use of Machine Learning techniques to counter the spread of these counterfeit samples.
Through several ML models, we are able to formalize an authentication method that is resilient against several attacks and scalable to many different devices.
In particular, we focused on the use of authentication and identification mechanisms that can rapidly and efficiently detect the legitimacy of a battery.
We proposed two different methodologies, \textbf{DCAuth} and \textbf{EISthentication}, which respectively use Differential Capacity Analysis data and Electrochemical Impedance Spectroscopy data to perform authentication on each battery cell.
The data is retrievable through the regular usage of the battery, and thus our techniques do not rely on any external devices.
Our models are able to detect both battery Li-ion architectures and battery models with high accuracy, obtaining F1 scores that reach up to 0.94 with the best classifier (i.e., Random Forest in battery model authentication).

\paragraph{Future Works}
As stated in Section~\ref{subsec:collection}, all the datasets used in this paper have been found in the literature.
While the overall number of battery models and architectures is still relevant, real-world implementations might consider several more.
Unfortunately, collecting this type of data is often expensive and requires specific expertise.
We believe that our methodologies can have a great impact on the battery industry and can benefit user safety; thus, we invite researchers to integrate our results with their data, when available, and report their findings.
Through practical implementations of our methodologies, the counterfeit battery market can be opposed by ensuring the legitimacy of each battery cell and thus guaranteeing the safety of users.

\bibliographystyle{ACM-Reference-Format}
\balance
\bibliography{references}

\clearpage
\nobalance
\appendix

\section{Results Details}
\label{app:results}

In Section~\ref{sec:results}, we display several graphs to show the results of our evaluation.
In this Appendix section, we numerically report those results.

\subsection{DCAuth}
\label{subapp:dcauth}

First, we report the results for DCAuth.

\subsubsection{Identification}
\label{subsubapp:dcauth-ident}

We start by showing in Table~\ref{apptab:dcauthident} the results of the identification task for DCAuth for both architecture identification and battery model identification (i.e., results shown in Figure~\ref{fig:dcauth-ident}).
Additionally, we include also the values for accuracy, precision, and recall.

\begin{table}[!htpb]
  \centering
  \caption{Scores for DCAuth in the identification task.}
  \label{apptab:dcauthident}
  \begin{subtable}[h]{.45\textwidth}
  \centering
  \caption{Architecture identification.}
  \label{subapptab:dcauthident1}
  \begin{tabular}{lcccc}
    \toprule
    \textbf{Model} & \textbf{Accuracy} & \textbf{Precision} & \textbf{Recall} & \textbf{F1 Score} \\
    \midrule
    AB & 0.407 & 0.267 & 0.403 & 0.242 \\
    DT & 0.994 & 0.994 & 0.994 & 0.994 \\
    GNB & 0.207 & 0.596 & 0.205 & 0.085 \\
    KNN & 0.948 & 0.949 & 0.948 & 0.948 \\
    NN & 0.965 & 0.965 & 0.965 & 0.964 \\
    QDA & 0.971 & 0.972 & 0.971 & 0.970 \\
    RF & 0.998 & 0.998 & 0.998 & 0.998 \\
    SVM & 0.978 & 0.978 & 0.978 & 0.978 \\
    \bottomrule
    \multicolumn{5}{l}{}
  \end{tabular}
  \end{subtable}
  \hfill
  \begin{subtable}[h]{.45\textwidth}
  \centering
  \caption{Battery model identification.}
  \label{subapptab:dcauthident2}
  \begin{tabular}{lcccc}
    \toprule
    \textbf{Model} & \textbf{Accuracy} & \textbf{Precision} & \textbf{Recall} & \textbf{F1 Score} \\
    \midrule
    AB & 0.198 & 0.123 & 0.242 & 0.139 \\
    DT & 0.802 & 0.835 & 0.827 & 0.800 \\
    GNB & 0.189 & 0.294 & 0.254 & 0.215 \\
    KNN & 0.802 & 0.812 & 0.817 & 0.797 \\
    NN & 0.764 & 0.779 & 0.784 & 0.766 \\
    QDA & 0.283 & 0.458 & 0.298 & 0.274 \\
    RF & 0.896 & 0.909 & 0.914 & 0.899 \\
    SVM & 0.774 & 0.781 & 0.785 & 0.768 \\
    \bottomrule
  \end{tabular}
  \end{subtable}
\end{table}

We also show in Table~\ref{apptab:dcauthident-sep} the explicit F1 scores obtained with the models when considering separate charge and discharge cycles (i.e., results shown in Figure~\ref{fig:dcauth-ident-sep}).
In this Table, we consider \textit{Arc} as architecture identification and \textit{BM} as battery model identification.
Subscript \textit{C} refers to charging cycle data, and subscript \textit{D} refers to discharging cycle data. 

\begin{table}[!htpb]
  \centering
  \caption{F1 Scores for DCAuth in the identification task when considering separate charging and discharging cycles.}
  \label{apptab:dcauthident-sep}
  \begin{tabular}{lcccc}
    \toprule
    \textbf{Model} & \textbf{Arc$_{C}$} & \textbf{Arc$_{D}$} & \textbf{BM$_{C}$} & \textbf{BM$_{D}$} \\
    \midrule
    AB & 0.081 & 0.309 & 0.236 & 0.249 \\
    DT & 0.986 & 0.992 & 0.811 & 0.863 \\
    GNB & 0.082 & 0.102 & 0.134 & 0.224 \\
    KNN & 0.928 & 0.936 & 0.818 & 0.799 \\
    NN & 0.956 & 0.976 & 0.612 & 0.830 \\
    QDA & 0.953 & 0.981 & 0.349 & 0.367 \\
    RF & 0.994 & 0.997 & 0.857 & 1 \\
    SVM & 0.972 & 0.985 & 0.798 & 0.849 \\
    \bottomrule
  \end{tabular}
\end{table}

\subsubsection{Authentication}
\label{subsubapp:dcauth-auth}

We show in Table~\ref{apptab:dcauthauth} the results of the authentication task for DCAuth for both architecture authentication and battery model authentication (i.e., results shown in Figure~\ref{fig:dcauth-auth}).
Additionally, we include also the values for accuracy, precision, and recall.

\begin{table}[!htpb]
  \centering
  \caption{Scores for DCAuth in the authentication task. Results are averaged with respect to the different balance levels for the dataset distribution.}
  \label{apptab:dcauthauth}
  \begin{subtable}[h]{.45\textwidth}
  \centering
  \caption{Architecture authentication.}
  \label{subapptab:dcauthauth1}
  \begin{tabular}{lcccc}
    \toprule
    \textbf{Model} & \textbf{Accuracy} & \textbf{Precision} & \textbf{Recall} & \textbf{F1 Score} \\
    \midrule
    AB & 0.998 & 0.998 & 0.996 & 0.997 \\
    DT & 0.993 & 0.991 & 0.992 & 0.991 \\
    GNB & 0.438 & 0.425 & 0.800 & 0.448 \\
    KNN & 0.967 & 0.962 & 0.953 & 0.957 \\
    NN & 0.976 & 0.973 & 0.966 & 0.969 \\
    QDA & 0.957 & 0.954 & 0.944 & 0.945 \\
    RF & 0.998 & 0.999 & 0.996 & 0.998 \\
    SVM & 0.976 & 0.968 & 0.971 & 0.969 \\
    \bottomrule
    \multicolumn{5}{l}{}
  \end{tabular}
  \end{subtable}
  \hfill
  \begin{subtable}[h]{.45\textwidth}
  \centering
  \caption{Battery model authentication.}
  \label{subapptab:dcauthauth2}
  \begin{tabular}{lcccc}
    \toprule
    \textbf{Model} & \textbf{Accuracy} & \textbf{Precision} & \textbf{Recall} & \textbf{F1 Score} \\
    \midrule
    AB & 0.960 & 0.942 & 0.930 & 0.933 \\
    DT & 0.930 & 0.877 & 0.902 & 0.885 \\
    GNB & 0.680 & 0.308 & 0.496 & 0.350 \\
    KNN & 0.919 & 0.867 & 0.872 & 0.864 \\
    NN & 0.914 & 0.874 & 0.846 & 0.848 \\
    QDA & 0.640 & 0.535 & 0.722 & 0.531 \\
    RF & 0.964 & 0.958 & 0.924 & 0.937 \\
    SVM & 0.871 & 0.772 & 0.838 & 0.799 \\
    \bottomrule
  \end{tabular}
  \end{subtable}
\end{table}

We also show in Table~\ref{apptab:dcauthauth-sep} the explicit F1 scores obtained with the models when considering separate charge and discharge cycles.
Also in this Table, we consider \textit{Arc} as architecture identification and \textit{BM} as battery model identification.
Subscript \textit{C} refers to charging cycle data, and subscript \textit{D} refers to discharging cycle data. 

\begin{table}[!htpb]
  \centering
  \caption{F1 Scores for DCAuth in the authentication task when considering separate charging and discharging cycles.}
  \label{apptab:dcauthauth-sep}
  \begin{tabular}{lcccc}
    \toprule
    \textbf{Model} & \textbf{Arc$_{C}$} & \textbf{Arc$_{D}$} & \textbf{BM$_{C}$} & \textbf{BM$_{D}$} \\
    \midrule
    AB & 0.994 & 0.998 & 0.887 & 0.929 \\
    DT & 0.988 & 0.992 & 0.872 & 0.89 \\
    GNB & 0.456 & 0.255 & 0.354 & 0.413 \\
    KNN & 0.955 & 0.968 & 0.776 & 0.867 \\
    NN & 0.973 & 0.973 & 0.824 & 0.87 \\
    QDA & 0.871 & 0.95 & 0.54 & 0.452 \\
    RF & 0.995 & 0.997 & 0.905 & 0.95 \\
    SVM & 0.974 & 0.984 & 0.851 & 0.896 \\
    \bottomrule
  \end{tabular}
\end{table}

Finally, in Table~\ref{apptab:dcauthauth-balance}, we show the explicit values for the False Acceptance Rate and False Rejection Rate at the varying of the balance level of the labels in the dataset distribution (i.e., results shown in Figure~\ref{fig:dcauth-auth-farfrr}).

\begin{table}[!htpb]
  \centering
  \caption{FAR and FRR for DCAuth for each dataset distribution. Results are averaged with respect to architecture authentication and battery model authentication.}
  \label{apptab:dcauthauth-balance}
  \begin{subtable}[h]{.45\textwidth}
  \centering
  \caption{False Acceptance Rate.}
  \label{subapptab:dcauthauthfar1}
  \begin{tabular}{lcccc}
    \toprule
    \textbf{Model} & \textbf{50/50} & \textbf{60/40} & \textbf{70/30} & \textbf{80/20} \\
    \midrule
    AB & 0.015 & 0.038 & 0.044 & 0.022 \\
    DT & 0.066 & 0.058 & 0.060 & 0.079 \\
    GNB & 0.475 & 0.510 & 0.270 & 0.270 \\
    KNN & 0.074 & 0.090 & 0.099 & 0.080 \\
    NN & 0.094 & 0.066 & 0.085 & 0.062 \\
    QDA & 0.174 & 0.187 & 0.294 & 0.324 \\
    RF & 0.027 & 0.016 & 0.023 & 0.020 \\
    SVM & 0.118 & 0.122 & 0.153 & 0.126 \\
    \bottomrule
    \multicolumn{5}{l}{}
  \end{tabular}
  \end{subtable}
  \hfill
  \begin{subtable}[h]{.45\textwidth}
  \centering
  \caption{False Rejection Rate.}
  \label{subapptab:dcauthauthfar2}
  \begin{tabular}{lcccc}
    \toprule
    \textbf{Model} & \textbf{50/50} & \textbf{60/40} & \textbf{70/30} & \textbf{80/20} \\
    \midrule
    AB & 0.026 & 0.016 & 0.007 & 0.016 \\
    DT & 0.046 & 0.010 & 0.012 & 0.026 \\
    GNB & 0.090 & 0.080 & 0.137 & 0.132 \\
    KNN & 0.050 & 0.054 & 0.026 & 0.040 \\
    NN & 0.054 & 0.038 & 0.028 & 0.044 \\
    QDA & 0.156 & 0.124 & 0.034 & 0.034 \\
    RF & 0.028 & 0.016 & 0.005 & 0.020 \\
    SVM & 0.071 & 0.049 & 0.028 & 0.042 \\
    \bottomrule
  \end{tabular}
  \end{subtable}
\end{table}

\subsection{EISthentication}
\label{subapp:eisthentication}

We now report results for EISthentication.

\subsubsection{Identification}
\label{subsubapp:eis-ident}

We start by showing in Table~\ref{apptab:eisident} the results of the identification task for DCAuth for both architecture identification and battery model identification (i.e., results shown in Figure~\ref{fig:eis-ident}).
Additionally, we include also the values for accuracy, precision, and recall.

\begin{table}[!htpb]
  \centering
  \caption{Scores for EISthentication in the identification task.}
  \label{apptab:eisident}
  \begin{subtable}[h]{.45\textwidth}
  \centering
  \caption{Architecture identification.}
  \label{subapptab:eisident1}
  \begin{tabular}{lcccc}
    \toprule
    \textbf{Model} & \textbf{Accuracy} & \textbf{Precision} & \textbf{Recall} & \textbf{F1 Score} \\
    \midrule
    AB & 0.667 & 0.609 & 0.714 & 0.633 \\
    DT & 0.963 & 0.969 & 0.964 & 0.964 \\
    GNB & 0.926 & 0.923 & 0.929 & 0.922 \\
    KNN & 0.852 & 0.860 & 0.866 & 0.856 \\
    NN & 0.815 & 0.825 & 0.821 & 0.809 \\
    QDA & 0.741 & 0.883 & 0.707 & 0.729 \\
    RF & 0.963 & 0.969 & 0.964 & 0.964 \\
    SVM & 0.704 & 0.709 & 0.713 & 0.710 \\
    \bottomrule
    \multicolumn{5}{l}{}
  \end{tabular}
  \end{subtable}
  \hfill
  \begin{subtable}[h]{.45\textwidth}
  \centering
  \caption{Battery model identification.}
  \label{subapptab:eisident2}
  \begin{tabular}{lcccc}
    \toprule
    \textbf{Model} & \textbf{Accuracy} & \textbf{Precision} & \textbf{Recall} & \textbf{F1 Score} \\
    \midrule
    AB & 0.242 & 0.346 & 0.458 & 0.333 \\
    DT & 0.636 & 0.767 & 0.706 & 0.721 \\
    GNB & 0.242 & 0.285 & 0.326 & 0.267 \\
    KNN & 0.333 & 0.405 & 0.343 & 0.360 \\
    NN & 0.364 & 0.427 & 0.387 & 0.380 \\
    QDA & 0.273 & 0.367 & 0.279 & 0.281 \\
    RF & 0.818 & 0.879 & 0.892 & 0.883 \\
    SVM & 0.394 & 0.451 & 0.433 & 0.422 \\
    \bottomrule
  \end{tabular}
  \end{subtable}
\end{table}

\subsubsection{Authentication}
\label{subsubapp:eis-auth}

We show in Table~\ref{apptab:eisauth} the results of the authentication task for DCAuth for both architecture authentication and battery model authentication (i.e., results shown in Figure~\ref{fig:eis-auth}).
Additionally, we include also the values for accuracy, precision, and recall.

\begin{table}[!htpb]
  \centering
  \caption{Scores for EISthentication in the authentication task. Results are averaged with respect to the different balance levels for the dataset distribution.}
  \label{apptab:eisauth}
  \begin{subtable}[h]{.45\textwidth}
  \centering
  \caption{Architecture authentication.}
  \label{subapptab:eisauth1}
  \begin{tabular}{lcccc}
    \toprule
    \textbf{Model} & \textbf{Accuracy} & \textbf{Precision} & \textbf{Recall} & \textbf{F1 Score} \\
    \midrule
    AB & 0.965 & 0.914 & 0.968 & 0.926 \\
    DT & 0.967 & 0.921 & 0.966 & 0.932 \\
    GNB & 0.954 & 0.891 & 0.977 & 0.910 \\
    KNN & 0.945 & 0.961 & 0.841 & 0.882 \\
    NN & 0.918 & 0.881 & 0.829 & 0.828 \\
    QDA & 0.860 & 0.794 & 0.752 & 0.730 \\
    RF & 0.979 & 0.960 & 0.977 & 0.962 \\
    SVM & 0.850 & 0.680 & 0.665 & 0.654 \\
    \bottomrule
    \multicolumn{5}{l}{}
  \end{tabular}
  \end{subtable}
  \hfill
  \begin{subtable}[h]{.45\textwidth}
  \centering
  \caption{Battery model authentication.}
  \label{subapptab:eisauth2}
  \begin{tabular}{lcccc}
    \toprule
    \textbf{Model} & \textbf{Accuracy} & \textbf{Precision} & \textbf{Recall} & \textbf{F1 Score} \\
    \midrule
    AB & 0.898 & 0.877 & 0.823 & 0.834 \\
    DT & 0.852 & 0.796 & 0.792 & 0.767 \\
    GNB & 0.737 & 0.478 & 0.628 & 0.520 \\
    KNN & 0.724 & 0.505 & 0.510 & 0.492 \\
    NN & 0.727 & 0.584 & 0.580 & 0.566 \\
    QDA & 0.702 & 0.580 & 0.653 & 0.586 \\
    RF & 0.914 & 0.850 & 0.843 & 0.838 \\
    SVM & 0.704 & 0.317 & 0.378 & 0.338 \\
    \bottomrule
  \end{tabular}
  \end{subtable}
\end{table}

Finally, in Table~\ref{apptab:eisauth-balance}, we show the explicit values for the False Acceptance Rate and False Rejection Rate at the varying of the balance level of the labels in the dataset distribution (i.e., results shown in Figure~\ref{fig:eis-auth-farfrr}).

\begin{table}[!htpb]
  \centering
  \caption{FAR and FRR for EISthentication for each dataset distribution. Results are averaged with respect to architecture authentication and battery model authentication.}
  \label{apptab:eisauth-balance}
  \begin{subtable}[h]{.45\textwidth}
  \centering
  \caption{False Acceptance Rate.}
  \label{subapptab:eisauthfar1}
  \begin{tabular}{lcccc}
    \toprule
    \textbf{Model} & \textbf{50/50} & \textbf{60/40} & \textbf{70/30} & \textbf{80/20} \\
    \midrule
    AB & 0.044 & 0.092 & 0.055 & 0.226 \\
    DT & 0.100 & 0.073 & 0.166 & 0.227 \\
    GNB & 0.185 & 0.188 & 0.058 & 0.164 \\
    KNN & 0.200 & 0.163 & 0.223 & 0.314 \\
    NN & 0.153 & 0.316 & 0.272 & 0.329 \\
    QDA & 0.163 & 0.288 & 0.416 & 0.386 \\
    RF & 0.050 & 0.080 & 0.062 & 0.104 \\
    SVM & 0.126 & 0.233 & 0.286 & 0.234 \\
    \bottomrule
    \multicolumn{5}{l}{}
  \end{tabular}
  \end{subtable}
  \hfill
  \begin{subtable}[h]{.45\textwidth}
  \centering
  \caption{False Rejection Rate.}
  \label{subapptab:eisauthfar2}
  \begin{tabular}{lcccc}
    \toprule
    \textbf{Model} & \textbf{50/50} & \textbf{60/40} & \textbf{70/30} & \textbf{80/20} \\
    \midrule
    AB & 0.106 & 0.074 & 0.017 & 0.040 \\
    DT & 0.116 & 0.103 & 0.021 & 0.046 \\
    GNB & 0.036 & 0.030 & 0.080 & 0.082 \\
    KNN & 0.242 & 0.188 & 0.098 & 0.092 \\
    NN & 0.210 & 0.208 & 0.092 & 0.076 \\
    QDA & 0.240 & 0.138 & 0.112 & 0.089 \\
    RF & 0.052 & 0.063 & 0.006 & 0.050 \\
    SVM & 0.320 & 0.188 & 0.144 & 0.115 \\
    \bottomrule
  \end{tabular}
  \end{subtable}
\end{table}

\subsection{Time and Space Overhead}
\label{subapp:complexity}

One of the advantages of both DCAuth and EISthentication is their compatibility with most devices.
Indeed, as anticipated in Section~\ref{subsec:dca} and Section~\ref{subsec:eis}, both DCA and EIS data can be retrieved from almost any battery-powered system.
Also, in Section~\ref{subsec:models}, we foresaw the potential mobile implementation of our tools and thus did not include any computationally-heavy Deep Learning algorithms in our model list.
While processing time might differ depending on various factors (e.g., size of the test set, computational power, CPU capabilities), once trained, the models have a fixed size which affects computational time.
Furthermore, a practical implementation of the system might be designed in such a way in which model training is performed outside the device and data retrieval and testing are performed on the device.
Through updates, manufacturers can enlarge their identification/authentication capabilities to include more battery models or architectures.
Thus, in Table~\ref{tab:complexity}, we show the size of each model for both DCAuth and EISthentication.
We also include computational time estimated in our local machine (specifications are available in Appendix~\ref{app:specs}).
While highly dependent on the computational capabilities of the device, comparisons between model size and test timings show that there is no linear relation between the two.
Nonetheless, all models manage to keep a small model size and low computational overhead.

\begin{table}[!htpb]
  \centering
  \caption{Complexity.}
  \label{tab:complexity}
  \begin{tabular}{lcccc}
    \toprule
    \textbf{Model} & \textbf{Time$_{DCA}$} & \textbf{Size$_{DCA}$} & \textbf{Time$_{EIS}$} & \textbf{Size$_{EIS}$} \\
    \midrule
    AB & 15.492 ms & 75 kB & 8.523 ms & 59 kB \\
    DT & 3.892 ms & 31 kB & 2.881 ms & 20 kB \\
    GNB & 4.687 ms & 53 kB & 3.192 ms & 33 kB \\
    KNN & 12.951 ms & 4800 kB & 7.1 ms & 263 kB \\
    NN & 4.595 ms & 2600 kB & 3.204 ms & 1200 kB \\
    QDA & 7.856 ms & 3100 kB & 4.435 ms & 271 kB \\
    RF & 13.661 ms & 348 kB & 13.288 ms & 221 kB \\
    SVM & 9.854 ms & 500 kB & 2.99 ms & 158 kB \\
    \bottomrule
  \end{tabular}
\end{table}

\section{Hardware and Software Configuration}
\label{app:specs}

All experiments have been conducted on a workstation with the following configurations.
\begin{itemize}
    \item \textbf{CPU}: AMD Ryzen 5 3600X.
    \item \textbf{RAM}: 32 GB at 3200 MT/s
    \item \textbf{Operating System}: Ubuntu 20.04.4 LTS.
    \item \textbf{Software}: Python 3.8.10.
\end{itemize}
All models are implemented with the \texttt{Scikit-learn} Python package, which does not natively support GPU acceleration.
Thus, each processing step has been performed on the CPU.
Other Python packages and their relative versions can be found in the requirements file present in each of the repositories.

\end{document}